\documentclass[prd,superscriptaddress,nofootinbib,preprintnumbers]{revtex4}
\usepackage{amsmath,amsfonts,amssymb}
\usepackage{graphicx}
\usepackage{dcolumn}
\usepackage[mathlines]{lineno}
\usepackage{float}
\usepackage{hyperref}
\usepackage[usenames,dvipsnames]{color}
\usepackage{etex}
\usepackage{booktabs}
\usepackage{adjustbox}

\begin{document}

\title{Nonholomorphic $A_4$ modular invariance for fermion masses and mixing in $SU(5)$ GUT}

\author{Mohamed Amin Loualidi}
\email{ma.loualidi@uaeu.ac.ae}
\affiliation{Department of physics, United Arab Emirates University, Al-Ain, UAE}

\author{Mohamed Miskaoui}
\email{m.miskaoui@gmail.com}
\affiliation{Department of physics, Faculty of Science, Mohammed V University in Rabat, 10090 Rabat, Morocco}

\author{Salah Nasri}
\email{snasri@uaeu.ac.ae, salah.nasri@cern.ch}
\affiliation{Department of physics, United Arab Emirates University, Al-Ain, UAE}

\begin{abstract}
Addressing the fermion flavor structures using modular invariance is a challenging task in the framework of quark-lepton unification. Building on recent applications of modular symmetry in nonsupersymmetric models, we propose the first renormalizable $SU(5)$ grand unified theory incorporating level 3 nonholomorphic modular symmetry, $\Gamma_3 \simeq A_4$. This framework constrains Yukawa couplings to polyharmonic Maa{\ss} forms, significantly reducing the number of free parameters while enhancing the predictive power of the models. We present a comprehensive analysis of fermion masses and mixing while tackling key grand unified theory (GUT) queries such as gauge coupling unification and proton decay. Beyond the minimal $SU(5)$ framework, the Higgs sector incorporates the $45_H$ dimensional Higgs field crucial in differentiating the masses of down quarks and charged leptons, and the fermion sector is extended with three right-handed neutrinos enabling neutrino masses via the type-I seesaw mechanism. We analyze two benchmark models with distinct modular weight and $A_4$ charge assignments. The predicted effective Majorana mass $m_{\beta \beta}$ values align with current neutrinoless double-beta decay experiments, and the effective neutrino mass $m_\beta$ is within the reach of future beta decay searches. The predicted sum of neutrino masses, $\sum m_i$, satisfies the upper bound set by recent cosmological observations. The gauge coupling unification is achieved through a light scalar triplet $\phi_3 \sim (3,3,-1/3)$ and a scalar octet $\phi_5 \sim (8,2,1/2)$ belonging to the $45_H$ Higgs, while proton decay constraints require that the contribution of the $45_H$ Higgs to the up-quark mass matrix remains highly suppressed.
\end{abstract}


\maketitle
\section{Introduction}
\label{introduction} 
The quest to understand the origins of fermion masses and their mixing remains a central challenge in particle physics. The distinct mass hierarchies and mixing behaviors observed in the lepton and quark sectors suggest that new physics may be at play, especially with the discovery of neutrino oscillations pointing to physics beyond the standard model (SM) \cite{Super, SNO}. In particular, unlike quarks where mixing angles are small, neutrino experiments have revealed that two angles in the Pontecorvo-Maki-Nakagawa-Sakata (PMNS) matrix are significantly large \cite{PDG}. Furthermore, while charge conjugation and parity ({\it CP}) symmetry violation is well established in the quark sector--first observed in kaon decays in the 1960s \cite{Christenson:1964fg}--the idea of {\it CP} violation in the lepton sector remains an intriguing possibility yet to be experimentally confirmed. Meanwhile, quark and charged lepton masses originate from their Yukawa interactions with the Higgs field, but the SM offers no explanation for the values of these couplings, leaving them as free parameters determined only by experiment. Additionally, the SM provides no underlying reason for the specific charge assignments of particles or the cancellation of anomalies. Given all these factors, and the fact that he renormalization group (RG) evolution of the three gauge couplings in the SM shows a tendency toward unification at a high energy scale, suggest the need for a more predictive and unified theory.
One approach to addressing these open questions is to explore symmetries beyond the gauge symmetry of the SM. In particular, non-Abelian discrete symmetries, which act on different generations of matter particles, naturally lead to large lepton mixing angles. This provides a structured framework for reproducing certain known lepton mixing patterns. For recent reviews on this topic, see Refs. \cite{King:2017guk, Xing:2020ijf, Feruglio:2019ybq, Ding:2024ozt}. Grand Unified Theories (GUTs) \cite{Pati:1973rp, Pati:1974yy, Georgi:1974sy, Georgi:1974yf, Georgi:1974my, Fritzsch:1974nn}, on the other hand, unify quarks and leptons within the same multiplets, offering an explanation for the charge quantization and anomaly cancellation while reducing the number of free parameters. 

\medskip

Although combining a non-Abelian symmetry with a GUT appears to be the ideal framework for addressing the shortcomings of the SM, both approaches come with their own challenges and complexities. In flavor model building, introducing a non-Abelian symmetry necessitates breaking it at some stage, which requires additional degrees of freedom in the form of scalar fields known as flavons. These flavons dictate the dynamics of symmetry breaking by requiring intricate vacuum alignments along different directions in flavor space across different fermion sectors. Thus, constructing a viable flavon potential becomes highly complex, often involving the introduction of additional discrete Abelian groups to eliminate unwanted operators and ensure the correct vacuum alignment. This complexity in both symmetry breaking and flavon potential construction makes these models difficult to manage, ultimately limiting their simplicity and predictivity. On the other hand, GUTs in their minimal realizations inherently impose degeneracy among fermion masses across different sectors, conflicting with experimental observations. Furthermore, they fail to achieve gauge coupling unification (GCU) without violating proton decay bounds set by Super-Kamiokande (SK) \cite{Super-Kamiokande:2020wjk}. Focusing on the Georgi-Glashow $SU(5)$ model, which embeds the SM gauge group $G_{SM} = SU(3)_C \times SU(2)_L \times U(1)_Y$ within a simple Lie group, this GCU issue is naturally resolved in its supersymmetric extension [SUSY $SU(5)$] \cite{Sakai:1981gr, Dimopoulos:1981zb}. There, the contributions of superpartners to the RG running drive GCU at $\mathcal{O}(10^{16})~\text{GeV}$ \cite{Ellis:1990wk, Amaldi:1991cn, Langacker:1991an, Giunti:1991ta}. However, the lack of experimental confirmation of low-energy SUSY at current collider limits challenges its viability, motivating renewed interest in alternative scenarios, with non-supersymmetric GUTs as a promising alternative.

\medskip

To address the fermion flavor structure without the complications of flavon vacuum alignment, modular invariance has been proposed as a flavor symmetry, offering a promising direction for flavor model building \cite{Feruglio:2017spp}. This approach first emerged in specific string compactifications \cite{Ferrara:1989bc, Chun:1989se, Lauer:1990tm}, where Yukawa couplings are described as modular forms, which are holomorphic functions of a complex scalar field $\tau$, known as the modulus. SUSY naturally provides a framework for constructing such models, in which the modulus acquires a vacuum expectation value (VEV), serving as the sole source of flavor symmetry breaking. Consequently, the introduction of flavons becomes unnecessary, thereby eliminating the associated vacuum alignment problem. In this framework, the superpotential must be invariant under the modular group $\Gamma = SL(2, \mathbb{Z})$ or its projective counterpart, the inhomogeneous modular group $PSL(2, \mathbb{Z})$, defined as the quotient of $SL(2, \mathbb{Z})$ by its center.
Although these modular groups are infinite, considering the quotient of $PSL(2,\mathbb{Z})$ by its principal congruence subgroups, $\bar{\Gamma}(N)$, results in finite modular groups\footnote{An early effort to link ordinary non-Abelian discrete symmetries with the modular group in flavor model building was introduced in Ref. \cite{Altarelli:2005yx}, where an $A_4$ neutrino model was constructed, exploring the origin of the $A_4$ group as a subgroup of the modular group. Other early studies investigating known neutrino mixing patterns through the modular group $PSL(2,\mathbb{Z})$ were proposed in Refs. \cite{Luhn:2007sy, Luhn:2007yr}. One such approach fixed $\mathbb{Z} = 5$, showing that $PSL(2,5)$ is isomorphic to the icosahedral group $A_5$. Furthermore, the group $PSL(2,7)$ was studied, highlighting its connection to the non-Abelian $S_4$ group, which appears as one of its maximal subgroups} $\Gamma_N = PSL(2,\mathbb{Z}) / \bar{\Gamma}(N)$ \cite{deAdelhartToorop:2011re}. For $N \leq 5$, these finite modular groups correspond to the well-known non-Abelian discrete symmetries: $\Gamma_2 \simeq S_3$, $\Gamma_3 \simeq A_4$, $\Gamma_4 \simeq S_4$, and $\Gamma_5 \simeq A_5$. Accordingly, both matter fields and the Yukawa couplings transform as irreducible representations of $\Gamma_N$. For a comprehensive overview of modular flavor symmetries, see Refs. \cite{Kobayashi:2023zzc, Ding:2023htn}.
As mentioned earlier, low-energy SUSY remains beyond experimental reach, making it imperative to investigate alternative frameworks such as modular invariance without SUSY. A significant breakthrough in this regard was presented in Ref. \cite{Qu:2024rns}, where modular flavor symmetry was successfully integrated into the lepton sector in a non-SUSY setting. The groundwork for this approach was laid in Ref. \cite{Ding:2020zxw}, where the authors proposed that automorphic forms could enable model-building independent of SUSY by replacing the holomorphicity constraint with a harmonic condition. This led to the replacement of Yukawa couplings, traditionally represented as modular forms, with polyharmonic Maa{\ss} forms, which incorporate a nonholomorphic component. As a result, these structures now serve as the fundamental mathematical framework governing flavor symmetries in non-SUSY models. So far, only a few studies have applied this novel approach to flavor model building \cite{Nomura:2024atp, Ding:2024inn, Li:2024svh, Nomura:2024nwh, Okada:2025jjo, Kobayashi:2025hnc}. However, no research has explored its application to quark-lepton unification, which remains a challenging task within the framework of modular symmetry without SUSY.

\medskip

In this paper, we present the first modular-invariant $\Gamma_3 \cong A_4$ flavor model within a non-SUSY\footnote{In the framework of the ordinary SUSY $SU(5)$ GUT in four dimensions, flavor models have been developed using various conventional non-Abelian symmetry groups, including $S_4$ \cite{Ishimori:2008fi, Hagedorn:2010th, Hagedorn:2012ut, Dimou:2015yng, Dimou:2015cmw} and its subgroups $A_4$ \cite{Altarelli:2008bg, Ciafaloni:2009ub, Ciafaloni:2009qs, Cooper:2010ik, Cooper:2012wf, Bjorkeroth:2015ora, Laamara:2018zpo}, $S_3$ \cite{Ramos:2024gcr}, and $D_4$ \cite{AhlLaamara:2017gsd, Miskaoui:2021nbm, Loualidi:2022dpx}. In addition, modular symmetries have been explored in the context of $\Gamma_4$ \cite{Ding:2021zbg, King:2021fhl, deMedeirosVarzielas:2023ujt}, $\Gamma_3$ \cite{deAnda:2018ecu, Chen:2021zty}, and $\Gamma_2$ \cite{Kobayashi:2019rzp, Du:2020ylx}. This list is not exhaustive, as it does not account for variations of the $SU(5)$ model or extensions to larger symmetry groups beyond $S_4$.} $SU(5)$ GUT. We investigate its implications for fermion masses and mixing in both the quark and lepton sectors, addressing key challenges of the minimal non-SUSY $SU(5)$ model, such as the failure of GCU and the issue of identical Yukawa couplings for down-type quarks and charged leptons. To resolve these limitations, we extend the Higgs sector by introducing a $45$-dimensional Higgs representation, enabling the differentiation of down-type quark and charged lepton masses. Additionally, we incorporate three right-handed (RH) neutrinos in the fermion sector to generate neutrino masses via the type-I seesaw mechanism \cite{Min, Yan, Gell, Glas, Moha}. 
Given the broad flexibility in assigning $A_4$ representations and modular weights to different fields, we impose theoretical constraints to systematically reduce the model space while maintaining consistency and achieving an optimal fit to experimental data. To simplify the scalar sector, all scalar multiplets are assumed to transform trivially under $A_4$ with zero modular weight. We further restrict our analysis to level-3 polyharmonic Maa{\ss} forms with even modular weights in the range $-4 \leq k_Y \leq 6$. Within this framework, we propose two benchmark renormalizable models, distinguished by the $A_4$ assignments and modular weights of the RH neutrinos and the three generations of matter fields in the conjugate $\overline{5}_{F_i}$ representations. In contrast, the $10_{F_i}$ matter fields share identical $A_4$ and modular weight charges in both models. Through a numerical analysis, we determine the best-fit values of the free parameters, including the modulus $\tau$, ensuring consistency with experimental observables in both the quark and lepton sectors. Our results indicate that the normal mass ordering (NO) is favored over the inverted ordering (IO) in benchmark model I, while both mass orderings remain viable in benchmark model II. We provide predictions for key neutrino parameters, including the sum of neutrino masses $\sum m_i$, the effective neutrino mass $m_\beta$, the Majorana effective mass $m_{\beta\beta}$, and the {\it CP}-violating phases. Notably, the atmospheric mixing angle aligns with experimental data only in the NO scenario, with model I favoring the higher octant and model II the lower octant. The predicted $m_{\beta\beta}$ values agree with current neutrinoless double-beta decay constraints, while $m_\beta$ remains below present experimental sensitivity but could be tested in future beta decay experiments. Similarly, the predicted sum of neutrino masses, $\sum m_i$, satisfies the upper bound set by recent cosmological observations. In the quark sector, the predicted mass ratios and mixing angles at the GUT scale are consistent with experimental data across all models. Moreover, we investigated the role of the $45_H$ Higgs representation in realizing GCU through threshold corrections induced by its light scalar components. In a scenario where the light spectrum contains a color-triplet scalar $\phi_3$ and a color-octet scalar $\phi_5$, unification must occur at a sufficiently high scale to suppress proton decay mediated by gauge bosons. Meanwhile, proton decay via the light Higgs triplet $\phi_3$ is further suppressed by stringent constraints on polyharmonic Maa{\ss} forms, whereas decays involving heavier scalar triplets remains negligible due to their GUT-scale masses.

\medskip

The rest of this paper is organized as follows. Sec. \ref{sec2} provides a concise introduction to modular invariance, modular forms, and polyharmonic Maa{\ss} forms. In Sec. \ref{sec3}, we present two benchmark $SU(5)$ models based on the finite modular group $ \Gamma_3 \cong A_4$ and derive the corresponding fermion Yukawa matrices. Sec. \ref{sec4} presents a numerical analysis of these benchmark models for both NO and IO neutrino mass spectra, including predictions for fermion mass ratios, mixing parameters, and key neutrino observables such as $ m_{\beta\beta}$, $ m_{\beta} $, and $ \sum{m_i} $. In Sec. \ref{sec5}, we explore the conditions for the GCU and consider scenarios where the components of the $45$ Higgs field exhibit a hierarchical mass spectrum. Additionally, we provide a brief analysis of the contribution of the scalar Higgs triplet from the $45$ Higgs field to proton decay. We conclude in Sec. \ref{sec6}. Additional details on the $A_4$ group and higher-weight polyharmonic Maa{\ss} forms at level 3 are provided in the appendix.
\section{Nonholomorphic modular symmetry}
\label{sec2} 
Flavor models based on modular invariance provide a compelling framework for understanding the flavor structure of fermions. In particular, $\mathcal{N}=1$ supersymmetric modular-invariant models within a bottom-up approach, have been remarkably successful in predicting lepton masses and mixing with only a few free parameters \cite{Feruglio:2017spp}. A key feature of these models is that matter fields in the superpotential transform under the inhomogeneous (or homogeneous) modular group $\Bar{\Gamma} = PSL(2,\mathbb{Z})$ ($\Gamma = SL(2,\mathbb{Z})$), which requires that the elements of the Yukawa coupling matrices are modular forms. These forms are holomorphic functions of the modulus $\tau$ in the upper half-plane $\mathbb{H}$, and they transform in specific ways under discrete subgroups of $SL(2, \mathbb{R})$ such as $\Bar{\Gamma}$ and $\Gamma$. Specifically, $\Gamma$ is the group of linear fractional transformations acting on $\tau$ as follows
\begin{equation}
    \tau \rightarrow \gamma \tau = \frac{a \tau + b}{c \tau + d}, \mkern9mu\text{where} \mkern9mu \gamma = \begin{pmatrix}
        a & b \\
        c & d
    \end{pmatrix}\in \Gamma \mkern9mu \text{with} \mkern9mu ad - bc = 1.
    \label{gt}
\end{equation}
Notice that the matrices $\pm \gamma$ act in the same way on $\tau$, so it is also common to work with the group\footnote{The difference between $\Gamma$ and $\Bar{\Gamma}$ lies in their structure: $\Gamma$ is the full modular group, consisting of $2 \times 2$ matrices with integer coefficients and unit determinant, while $\Bar{\Gamma}$ is the projective special linear group, defined as the quotient of $\Gamma$ by its center: $\Bar{\Gamma} = \Gamma / \{\pm I_2\}$, where $I_2$ is the identity matrix. This distinction reflects the removal of the overall sign ambiguity in $\Gamma$.} $PSL(2,\mathbb{Z})$. This group is infinite, noncompact, and  is generated by two elements, $S$ and $T$, given by
\begin{equation}
S = \begin{pmatrix} 0 & 1 \\ -1 & 0 \end{pmatrix}, \quad T = \begin{pmatrix} 1 & 1 \\ 0 & 1 \end{pmatrix},
\end{equation}
which satisfy the relations $S^2 = (ST)^3 = 1$. Under the action of these generators, $S$ transforms $\tau$ as $\tau \to \frac{-1}{\tau}$, whereas $T$ induces the shift $\tau \to 1+\tau$. 
\newline

In flavor models with modular invariance, finite modular groups with finite-dimensional representations play a fundamental role in constructing a well-defined flavor structure while keeping the number of parameters under control. To achieve this, chiral superfields and modular forms appearing in the superpotential are assumed to transform under representations of discrete finite modular groups, which are defined as $\Gamma_N = \Bar{\Gamma}/\Bar{\Gamma}(N)$ for the inhomogeneous modular group and $\Gamma_N^{\prime} = \Gamma/\Gamma(N)$ for the homogeneous one, where $N$ represents the level of the group. Here, $\Gamma(N)$ denotes the infinite principal congruence subgroups of $SL(2,\mathbb{Z})$. These subgroups act on $\tau$ similarly to $\Gamma$, but with additional modular congruence conditions defined as 
\begin{equation}
    \Gamma(N) = \left\{ \begin{pmatrix} a & b \\ c & d \end{pmatrix} \in SL(2, \mathbb{Z}) \Bigg| \begin{pmatrix} a & b \\ c & d \end{pmatrix} \equiv \begin{pmatrix} 1 & 0 \\ 0 & 1 \end{pmatrix} \mod N \right\}.
\end{equation}
For $N=2,3,4,5$, the finite modular groups $\Gamma_N$ are isomorphic to the permutation groups $S_3$, $A_4$, $S_4$ and $A_5$ \cite{deAdelhartToorop:2011re}, while the groups $\Gamma_N^{\prime}$ correspond to the double covers of the respective permutation groups \cite{Liu:2019khw}. Given how $\gamma$ acts on $\tau$ in Eq. \ref{gt}, a modular form $f(\tau)$ of level $N$ transforms under $\Gamma(N)$ as $f(\gamma \tau) = (c\tau + d)^{k} f(\tau)$ where $k$ is a positive integer called the modular weight. For a given weight $k$ and level $N$, the modular forms constitute a finite-dimensional vector space, $M_k(\Gamma(N))$, with dimension $k+1$. It can be readily shown that this space admits a basis in which a multiplet of modular forms $f_i(\tau)$ transforms according to a unitary representation $\rho$ of the finite group $\Gamma_N$ \cite{Feruglio:2017spp} as
\begin{equation}
    f_i(\gamma \tau) = (c\tau + d)^k \rho_{ij}(\gamma) f_j(\tau), \quad \gamma \in \Gamma_N,
\end{equation}
up to the automorhpy factor $(c\tau + d)^k$. For example, consider a trilinear term in the superpotential of the form ${\cal{W}}(\tau,\phi) \supset Y_{ijk}(\tau) \phi^i \phi^j \phi^k$. Here the modular forms $Y_{ijk}(\tau)$ transform in the representation $\rho_Y$ of a finite modular group ($\Gamma_N$ or $\Gamma_N^{\prime}$) and the superfields $\phi_i$ transform under the action of the modular group as follows
\begin{equation}
    Y_{ijk}(\tau) \xrightarrow{\gamma} Y_{ijk}(\gamma \tau) = (c \tau + d)^{k_Y} \rho_Y(\gamma) Y_{ijk}(\tau), \quad \phi_i \xrightarrow{\gamma} (c \tau + d)^{k_i} \rho_i(\gamma) \phi_i
    \label{yij}
\end{equation}
The term in $\cal{W}(\tau,\phi)$ is modular invariant if the modular weights and the representations in Eq. \ref{yij} satisfy: $k_Y = k_i + k_j + k_k$ and $\rho_Y \otimes \rho_i \otimes \rho_j \otimes \rho_k \supset \mathbf{1}$ where $\mathbf{1}$ is the trivial singlet of $\Gamma_N$ or $\Gamma_N^{\prime}$. Typically, the interplay between modular invariance and the holomorphic nature of the superpotential restricts the allowed terms, significantly reducing the number of free parameters and yielding highly predictive models. 

In the above formulation, SUSY ensures the holomorphic structure of Yukawa couplings as modular forms within the superpotential. However, the absence of experimental evidence for  low-energy SUSY casts doubt on its  existence, motivating the exploration of modular invariance in non-SUSY models, where the field content is significantly reduced. Indeed, a groundbreaking approach has been developed in Ref. \cite{Qu:2024rns} in which modular flavor symmetry was successfully applied to the lepton sector in non-SUSY context. This idea was first proposed in Ref. \cite{Ding:2020zxw}, where the authors suggested that the structure of automorphic forms\footnote{Automorphic forms generalize modular forms by extending their applicability to broader groups and spaces \cite{borel2007automorphic,borel1966algebraic}.} could offer a path to constructing models without relying on SUSY by replacing the holomorphicity constraint with the harmonic condition. For a single modulus $\tau = x + i y$, these automorphic forms coincide with harmonic Maa{\ss} forms $F$ which, unlike purely holomorphic modular functions, have a nonholomorphic component and satisfy the Laplace equation $\Delta_k F = 0$ where $\Delta_k$ is the weight $k$ hyperbolic Laplacian operator defined as
\begin{equation}
    \Delta_k = -y^2 \left(\frac{\partial^2}{\partial x^2} + \frac{\partial^2}{\partial y^2}\right) + iky \left( \frac{\partial}{\partial x} +i \frac{\partial}{\partial y} \right) = -4y^2 \frac{\partial}{\partial \tau}\frac{\partial}{\partial \bar{\tau}} + 2iky \frac{\partial}{\partial \bar{\tau}}.
\end{equation}
More generally, one can consider polyharmonic Maa{\ss} forms, which satisfy $\Delta_k^m F = 0$ for some positive integer $m$, with $m = 1$ corresponds to harmonic Maa{\ss} forms. Accordingly, polyharmonic Maa{\ss} forms of level $N$ have been attributed to the Yukawa couplings in the context of non-SUSY models, effectively replacing modular forms as the key mathematical objects governing flavor structures \cite{Qu:2024rns}. In this case, Yukawa couplings satisfy the same automorphy condition as modular forms, with the Laplacian condition replacing holomorphicity
\begin{eqnarray}
    Y(\gamma \tau) = (c\tau + d)^k Y(\tau), \quad  \Delta_k^m Y(\tau) = 0.
\end{eqnarray}
A key distinction from modular forms is that, in the case of polyharmonic Maa{\ss} forms, the weight $k$ is an even integer for $\Gamma_N$ and an integer for $\Gamma_N^{\prime}$. Moreover, these functions are required to grow at a controlled rate as $Im(\tau)\rightarrow \infty$ where they satisfy a moderate growth condition\cite{borel2007automorphic}: $Y(\tau) = {\cal{O}}(y^\alpha)$ for some $\alpha \in \cal{R}$ and which holds at all cusps of $\Gamma(N)$ \cite{lagarias2016polyharmonic}. When applied to flavor model building, the overall construction remains analogous to the modular forms case. At each given weight and level, polyharmonic Maa{\ss} forms form a finite-dimensional space, leading to constraints on the Yukawa interactions. Consequently, only a finite number of fermion mass terms are permitted, ensuring that modular invariance retains its predictive power. To illustrate, consider the inhomogeneous finite modular group $\Gamma_N$ and the Yukawa interaction: $-\mathcal{L} = Y_{ij}^{k_Y}(\tau) \bar{\Psi}_L^i \Phi \Psi_R^j + H.c.,$ where $\Psi_{L,R}$ are matter fields, and $\Phi$ denotes the Higgs field. Under a modular transformation $\gamma \in \bar{\Gamma}$, these fields transform as
\begin{equation}
    \bar{\Psi}_L \xrightarrow{\gamma} (c\tau + d)^{-k_{\bar{\Psi}}} \rho_{\bar{\Psi}}(\gamma) \bar{\Psi}_L, \quad \Psi_R \xrightarrow{\gamma} (c\tau + d)^{-k_{\Psi}} \rho_{\Psi}(\gamma) \Psi_R, \quad \Phi \xrightarrow{\gamma} (c\tau + d)^{-k_{\Phi}} \rho_{\Phi}(\gamma) \bar{\Psi}_L, 
\end{equation}
where $k_{\bar{\Psi}},~k_{\Psi},~k_{\Phi}$ are integer modular weights, and $\rho_{\bar{\Psi}},~\rho_{\Psi},~\rho_{\Phi}$ are irreducible representations of $\Gamma_N$. Modular invariance requires that the coupling $Y_{ij}(\tau)$ is a multiplet of even integer weight $k_Y$ and level $N$ polyharmonic Maa{\ss} forms that transforms in a representation $\rho_Y$ of $\Gamma_N$ as
\begin{equation}
     Y^{k_Y}(\tau) \xrightarrow{\gamma} Y^{k_Y}(\gamma \tau) = (c \tau + d)^{k_Y} \rho_Y(\gamma) Y^{k_Y}(\tau)
\end{equation}
where the modular weights and representations must satisfy $k_Y = k_{\bar{\Psi}} + k_{\Psi} + k_{\Phi}$ and  $\rho_Y \otimes \rho_{\bar{\Psi}} \otimes \rho_{\Psi} \otimes \rho_{\Phi} \supset \mathbf{1}$.

In this work, We focus on the level $N = 3$ polyharmonic Maa{\ss} forms which can be grouped into multiplets of $\Gamma_3 \simeq A_4$. These forms admit an expansion as a Fourier series given by \cite{Qu:2024rns}
\begin{equation}
    Y(\tau) = \sum_{n\in\frac{1}{N} \mathbb{Z}, n \geq 0} c^+(n)q^n + c^-(0) y^{1-k} + \sum_{n\in\frac{1}{N} \mathbb{Z}, n<0} c^-(n) \Gamma(1-k,-4\pi n y)q^n, \quad q \equiv e^{2 \pi i \tau}
\end{equation}
where the coefficients $c^\pm(n)$ and $c^\pm(0)$ are constants and $\Gamma(x,y)$ is the incomplete gamma function defined as
\begin{equation}
    \Gamma(x,y) = \int_y^{+\infty} e^{-t} t^{x-1} dt
\end{equation}
The expressions of the polyharmonic Maa{\ss} forms multiplets of level 3 are given in the Appendix.
\section{Implementing the nonholomorphic $A_4$ group in $SU(5)$ GUT}
\label{sec3}
In the minimal $SU(5)$ GUT model, the SM fermions are embedded into two irreducible representations:  $\bar{5}_{Fi}$ and $10_{Fi}$, where $i = 1,2,3$ labels the three generations. For a single generation, the fields in $\bar{5}_{F_1}$ consist of the charge-conjugate right-handed down-type quark $d^c$ and the left-handed lepton doublet $L = (\nu_L, e_L)$, while $10_{F_1}$ contains the charge-conjugate right-handed up-type quark $u^c$, the charge-conjugate right-handed charged lepton $e^c$, and the left-handed quark doublet $Q_L = (u_L, d_L)$. The Higgs sector includes a 24-dimensional adjoint Higgs field $H_{24}$, responsible for breaking $SU(5)$ down to the SM gauge group, and a fundamental five-dimensional Higgs field $5_H$, which contains both a color triplet scalar $T$ and the usual SM Higgs doublet $H$; $5_{H}\equiv H_{5} = (T,H)$. Omitting the $SU(5)$ and Lorentz indices, the Yukawa Lagrangian of the model is given by
\begin{equation}
\mathcal{L}_{Y} = Y_{\bar{5}}^{ij} 10_{F_i} \bar{5}_{F_j} 5_{H}^\ast + Y_{10}^{ij} 10_{F_i} 10_{F_j} 5_{H} + H.c.  
\label{ly}
\end{equation}
By decomposing these terms into their SM components, we identify $Y_{\bar{5}}^{ij}$ as encoding the Yukawa couplings for down-type quarks ($Y_d^{ij}$) and charged leptons ($Y_e^{ij}$), while $Y_{10}^{ij}$ governs the up-type quark Yukawa interactions ($Y_u^{ij}$). When the Higgs doublet acquires a VEV at the electroweak scale, it predicts the mass relations $m_e = m_d$, $m_\mu = m_s$, and $m_\tau = m_b$. While the third-generation relation is reasonable, those for the first two generations conflict with experimental data, where the observed mass ratios are $m_s/m_d \sim 20$ and $m_\mu/m_e \sim 200$ \cite{PDG}. Nearly fifty years ago, Georgi and Jarlskog \cite{GJ} proposed the first remedy for this issue. They demonstrated that the predicted mass relations could be altered through the application of group-theoretical Clebsch-Gordan (CG) factors, suggesting the need for a more abundant Higgs structure than what the minimal $SU(5)$ model offered. Specifically, they introduced an additional Higgs field, $45_{H}$, transforming in the 45-dimensional representation. This modification introduced an extra term in the Lagrangian given by $Y_{\bar{45}}^{ij} 10_{F_i} \bar{5}_{F_j} 45_{H}^\ast $, where the VEV of this new Higgs is expressed as
\begin{equation}
\left\langle (H_{45})_{\beta }^{\alpha 5}\right\rangle =\upsilon
_{45}(\delta _{\alpha }^{\beta }-4\delta _{4}^{\alpha }\delta _{\beta}^{4})\quad \text{with}\quad \alpha ,\beta =1,....,4  \label{v45}
\end{equation}
This structure plays a crucial role in differentiating between the masses of charged leptons and down quarks belonging to the first two generations. This distinction arises from the relative factor of $-3$ in the component $\left\langle (H_{45})_{5}^{45}\right\rangle = -3 \upsilon_{45}$. Beyond its role in fermion mass generation, the scalar components of $45_{H}$ can contribute to gauge coupling unification when some of them acquire masses below $M_{GUT}$. In such case, these light states remain active in the renormalization group equations (RGEs) at intermediate scales, affecting the beta functions and modifying the running of gauge couplings. Their contribution can lead to threshold corrections that shift the unification scale and improve the consistency of gauge coupling unification within the model.

We now analyze the modular-invariant $SU(5) \times A_4$ GUT in a renormalizable nonsupersymmetric framework, where the minimal $SU(5)$ model is extended with a $45_H$ Higgs, as motivated earlier, along with three RH neutrinos, $N_{1,2,3}^c$, to account for neutrino masses and mixing by employing the type-I seesaw mechanism. We introduce two benchmark models, each characterized by distinct modular weights and $A_4$ charge assignments for the various fields involved. In general, a renormalizable Yukawa Lagrangian that respects modular invariance can be formulated as
\begin{eqnarray}
    \mathcal{L}_Y^{\Gamma_3} &=& \sum_{(r \otimes r^{\prime})_p} \Bigg\{ 
    a_p \left[ (T_i T_j H_5)_r Y_{r'}^{(k_{T_i} + k_{T_j} + k_{H_5})}(\tau) \right]_\mathbf{1}  
    + b_p \left[ (T_i \bar{F}_j H_5^\ast)_r Y_{r'}^{(k_{T_i} +   k_{\bar{F}_j} + k_{H_5^\ast})}(\tau) \right]_\mathbf{1} \nonumber \\ &+& b_p^{\prime} \left[ (T_i \bar{F}_j H_{45}^\ast)_r Y_{r'}^{(k_{T_i} + k_{\bar{F}_j} + k_{H_{45}^\ast})}(\tau) \right]_\mathbf{1}  + c_p \left[ (\bar{F}_i N_j^c H_5)_r Y_{r'}^{(k_{\bar{F}_i} + k_{N_j^c} + k_{H_5})}(\tau) \right]_\mathbf{1}  \\
    &+& c_p^{\prime} \Lambda_p \left[ (N_i^c N_j^c)_r Y_{r'}^{(k_{N_i^c} + k_{N_j^c})}(\tau) \right]_\mathbf{1} 
    \Bigg\} + H.c., \nonumber
\end{eqnarray}
where the tensor products $(r \otimes r')_p$ encompass all possible combinations that yield the trivial singlet of $A_4$ [see Eq. \ref{tpr} in the Appendix]. The index $p = 1, 2, 3, \dots$ enumerates the distinct $A_4$-invariant coupling terms. For instance, consider an $A_4$ triplet $10_{F_i} \equiv T_i = (T_1, T_2, T_3)^T$ with weight 2 and an $A_4$ trivial singlet $5_H \equiv H_5$ with weight 0. For the first term in the above Yukawa Lagrangian, the number of invariant terms is determined by the possible weights for the polyharmonic Maa{\ss} forms that can be constructed. In this example, there are four such terms ($p = 1,2,3,4$) according to Eqs. \ref{y1m2}-\ref{y3m4} in the Appendix, given by
\begin{equation}
a_1 (T_i T_j H_5)_1 Y_1^{4}, \quad  
a_2 (T_i T_j H_5)_1'' Y_1'^{4}, \quad  
a_3 (T_i T_j H_5)_{3_S} Y_3^{4}, \quad  
a_4 (T_i T_j H_5)_{3_A} Y_3^{4}.  
\end{equation}
These correspond to the singlet and triplet contractions of the tensor product, ensuring that the resulting term remains invariant under $A_4$. Here, the antisymmetric contraction $(T_i T_j H_5)_{3_A}$ vanishes because the antisymmetric combination of two identical triplets is zero, and thus $a_4 = 0$. Based on the Lagrangian ${\cal{L}}_Y^{\Gamma_3}$ and the VEVs of $H_5$, given by $\langle H_5 \rangle = \upsilon_5$, and $H_{\bar{45}}$ in Eq. \ref{v45}, we derive the fermion mass matrices for up-type quarks, down-type quarks, charged leptons, and Dirac neutrinos after electroweak (EW) symmetry breaking, which can be expressed as
\begin{eqnarray}
    M_u = Y_{10} \upsilon_5, \quad M_d = Y_{\bar{5}} \upsilon_5^\ast + Y_{\bar{45}} \upsilon_{45}^\ast, \quad M_e = Y_{\bar{5}}^T \upsilon_5^\ast - 3 Y_{\bar{45}}^T \upsilon_{45}^\ast, \quad M_D = Y_D \upsilon_5
    \label{masses}
\end{eqnarray}
In the present framework, light neutrino masses are generated through the type-I seesaw mechanism. The mass matrix for the light neutrinos is expressed as $m_\nu = -M_D^T M_R^{-1} M_D$, where $M_R$ denotes the Majorana mass matrix, and $M_D$ is the Dirac mass matrix defined in Eq. \ref{masses}. For a more convenient numerical analysis, the Yukawa mass matrices can be reparametrized after diagonalizing the mass matrices, using the VEV of the SM Higgs doublet. Due to the presence of $45_H$, this VEV arises from the mixing of the Higgs doublets $H \in H_5$ and $H' \in H_{45}$, leading to $\upsilon^2 = \upsilon_5^2 + \upsilon_{45}^2 = (246~\text{GeV})^2$. Thus, we define the rescaled Yukawa matrices as
\begin{equation}
    \tilde{Y}_{10} = \frac{\upsilon_5}{\upsilon} Y_{10}, \quad 
    \tilde{Y}_{\bar{5}} = \frac{\upsilon_{\bar{5}}}{\upsilon} Y_{\bar{5}}, \quad 
    \tilde{Y}_{\bar{45}} = \frac{\upsilon_{\bar{45}}}{\upsilon} Y_{\bar{45}}, \quad 
    \tilde{Y}_D = \frac{\upsilon_5}{\upsilon} Y_{D},
\end{equation}
where the new Yukawa matrices remain proportional to the original ones, and the VEV ratios $\upsilon_5 / \upsilon$ and $\upsilon_{\bar{45}} / \upsilon$ can be absorbed into the coupling constants of each matrix. Accordingly, the mass matrices in Eq. \ref{masses} are rewritten as follows
\begin{equation}
        M_u = \tilde{Y}_{10} \upsilon, \quad M_d = (\tilde{Y}_{\bar{5}} + \tilde{Y}_{\bar{45}}) \upsilon, \quad M_e = (\tilde{Y}_{\bar{5}} - 3 \tilde{Y}_{\bar{45}})^T \upsilon, \quad M_D = \tilde{Y}_D \upsilon.
    \label{masses2}
\end{equation}
The flexibility in assigning $A_4$ representations and modular weights to different fields results in a variety of possible model constructions. To systematically reduce the number of models to explore, we impose specific constraints that maintain theoretical consistency while ensuring a good fit to experimental data in both the lepton and quark sectors. For simplicity, we assume a trivial scalar sector, where all scalar multiplets transform trivially under $A_4$ and carry zero modular weight. Additionally, we restrict our analysis to level-3 polyharmonic mass forms with even modular weights in the range $-4 \leq k_Y \leq 6$. Finally, we exclude two limiting cases: (i) all fermions assigned to $A_4$ singlets and (ii) all fermions assigned to $A_4$ triplets. The former effectively reduces to a simple $Z_3$ symmetry, undermining the motivation for flavor models with non-Abelian discrete groups, which naturally accommodate fermions in doublet and triplet representations for a structured description of neutrino mixing. The latter case, on the other hand, leads to an overconstrained model, making it significantly harder to reconcile with experimental data. With these assumptions in place, we now provide the details of two benchmark $SU(5) \times \Gamma_3$ models.

\medskip
\textbf{Model I:}
In this model, the three generations of matter fields in the $\mathbf{10}$ representation are assigned to the singlet irreducible representations of $A_4$ as $(T_1, T_2, T_3) \sim (1'', 1', 1)$, while those in the $\mathbf{\bar{5}}$ representation, and the RH neutrinos, transform as $A_4$ triplets.
\begin{table}[H]
\centering
\setlength{\tabcolsep}{8pt}
\begin{tabular}{|c||c|c|c|c|c|c|c|c|c|c|c|c|c|}
\hline
Model I & $T_1$ & $T_2$ & $T_3$ & $\bar{F}_i$ & $N_i^c$ & $H_5^\ast$ & $H_{45}^\ast$ & $Y_{\mathbf{1}}^{(-2)}$ & $Y_{\mathbf{3}}^{(-2)}$ & $Y_{\mathbf{3}}^{(0)}$ & $Y_{\mathbf{3}}^{(2)}$ & $Y_{\mathbf{1'}}^{(4)}$ & $Y_{\mathbf{1}}^{(6)}$  \\ \hline
$SU(5)$ & $10$ & $10$ & $10$ & $\bar{5}$ & $1$ & $\bar{5}$ & $\bar{45}$ & $1$ & $1$ & $1$ & $1$ & $1$ & $1$ \\ \hline
$A_4$ & $\mathbf{1''}$ & $\mathbf{1'}$ & $\mathbf{1}$ & $\mathbf{3}$ & $\mathbf{3}$ & $\mathbf{1}$ & $\mathbf{1}$ & $\mathbf{1}$ & $\mathbf{3}$ & $\mathbf{3}$ & $\mathbf{3}$ & $\mathbf{1'}$ & $\mathbf{1}$ \\ \hline
$k_I$ & $4$ & $2$ & $0$ & $-2$ & $0$ & $0$ & $0$ & $-2$ & $-2$ & $0$ & $2$ & $4$ & $6$ \\ \hline
\end{tabular}
\caption{Transformation properties of matter fields, Higgs fields, and polyharmonic Maa{\ss} forms in model I under $SU(5) \times A_4$ and their corresponding modular weights. The bold values here denotes the dimension of the irreducible representations under the $A_4$ symmetry (e.g., $\mathbf{1}$, $\mathbf{1'}$, $\mathbf{3}$), distinguishing them from the $SU(5)$ representations and modular weights (kI ).}
\label{tab1}
\end{table}
Table \ref{tab1} summarizes the $A_4$ charge assignments and the modular weights for all fields in the model, along with the polyharmonic Maa{\ss} forms ensuring the modular invariance of the Lagrangian. The renormalizable Yukawa interactions, preserving both $SU(5)$ gauge symmetry and the modular group $\Gamma_3$, are given by
\begin{eqnarray}
    -{\cal{L}}_Y^I &=& a_1 (T_1 T_2)_{\mathbf{1}} Y_\mathbf{1}^{(6)} H_5 + a_2 (T_1 T_3)_{\mathbf{1''}} Y_\mathbf{1'}^{(4)} H_5 + a_3 (T_2 T_2)_{\mathbf{1''}} Y_\mathbf{1'}^{(4)} H_5 + a_4 (T_3 T_3)_{\mathbf{1}} H_5 + b_1 (T_1 \bar{F}_j)_\mathbf{3} Y_\mathbf{3}^{(2)} H_5^\ast \nonumber \\
    &+& b_2 (T_2 \bar{F}_j)_\mathbf{3} Y_\mathbf{3}^{(0)} H_5^\ast + b_3 (T_3 \bar{F}_j)_\mathbf{3} Y_\mathbf{3}^{(-2)} H_5^\ast + b'_1 (T_1 \bar{F}_j)_\mathbf{3} Y_\mathbf{3}^{(2)} H_{45}^\ast + b'_2 (T_2 \bar{F}_j)_\mathbf{3} Y_\mathbf{3}^{(0)} H_{45}^\ast \\
    &+& b'_3 (T_3 \bar{F}_j)_\mathbf{3} Y_\mathbf{3}^{(-2)} H_{45}^\ast + c_1 (\bar{F}_i N_j^c)_\mathbf{1} Y_\mathbf{1}^{(-2)} H_5 + c_2 (\bar{F}_i N_j^c)_\mathbf{3} Y_\mathbf{3}^{(-2)} H_5 + c'_1 \Lambda_1 (N_i^c N_j^c)_\mathbf{1} + c'_2 \Lambda_2 (N_i^c N_j^c)_\mathbf{3} Y_\mathbf{3}^{(0)} + H.c. \nonumber
    \label{ly}
\end{eqnarray}
Following the decomposition of the tensor product of the irreducible representations of $A_4$ (see Appendix), the resulting structure of this Lagrangian gives rise to the following Yukawa matrices
\begin{eqnarray}
    Y_{10} &=& \begin{pmatrix}
        0 & a_1 Y_\mathbf{1}^{(6)} & a_2 Y_\mathbf{1'}^{(4)} \\
        a_1 Y_\mathbf{1}^{(6)} & a_3 Y_\mathbf{1'}^{(4)} & 0 \\
        a_2 Y_\mathbf{1'}^{(4)} & 0 & a_4
    \end{pmatrix}, \quad Y_{\bar{5}} = \begin{pmatrix}
        b_1 Y_{3,2}^{(2)} & b_2 Y_{3,3}^{(0)} & b_3 Y_{3,1}^{(-2)} \\ b_1 Y_{3,1}^{(2)} & b_2 Y_{3,2}^{(0)} & b_3 Y_{3,3}^{(-2)} \\
        b_1 Y_{3,3}^{(2)} & b_2 Y_{3,1}^{(0)} & b_3 Y_{3,2}^{(-2)}
    \end{pmatrix}, \nonumber \\ Y_{\bar{45}} &=& \begin{pmatrix}
        b'_1 Y_{3,2}^{(2)} & b'_2 Y_{3,3}^{(0)} & b'_3 Y_{3,1}^{(-2)} \\ b'_1 Y_{3,1}^{(2)} & b'_2 Y_{3,2}^{(0)} & b'_3 Y_{3,3}^{(-2)} \\
        b'_1 Y_{3,3}^{(2)} & b'_2 Y_{3,1}^{(0)} & b'_3 Y_{3,2}^{(-2)}
    \end{pmatrix}, \quad
    Y_D = \begin{pmatrix}
        c_1 Y_1^{(-2)} + 2 c_2 Y_{3,1}^{(-2)} & 0 & -2 c_2 Y_{3,2}^{(-2)} \\
        -2 c_2 Y_{3,3}^{(-2)} & 2 c_2 Y_{3,2}^{(-2)} & c_1  Y_1^{(-2)} \\ 0 & c_1 Y_1^{(-2)} - 2 c_2 Y_{3,1}^{(-2)} & 2 c_2 Y_{3,3}^{(-2)}
    \end{pmatrix}
    \label{ycou}
\end{eqnarray}
The fermion mass matrices are obtained using the same method employed to derive their expressions in Eq. \ref{tab:masses2}. We then factorize each Yukawa matrix with a coupling constant, allowing us to express the input parameters for our numerical study in terms of overall mass scales, coupling constant ratios, and the complex modulus $\tau$, which is associated with polyharmonic Maa{\ss} forms. The phases of the parameters $a_3,~b_1,~b_2,~b_3,~b'_1,~c_1,~c_2,~c'_1$ are unphysical, while the remaining parameters are complex. This model features three overall mass scales: $a_3 \upsilon$ in the up-quark mass matrix, $b_3 \upsilon$ in both the down-quark and charged lepton mass matrices, and an additional factor for the neutrino mass matrix, which is discussed below.
For the Majorana mass matrix of the RH neutrinos denoted as $M_R$, it is derived from the last two terms in Eq. \ref{ly}. Then, the lightest neutrino masses are generated by the type-I seesaw mechanism where $m_\nu = -M_D^T M_R^{-1} M_D$. Assuming for simplicity that $\Lambda_1 = \Lambda_2 = \Lambda$, the expressions for $M_R$ and $M_D$ for model I are given by
\begin{equation}
    M_R = c'_1 \Lambda \begin{pmatrix}
        1 + 2 \frac{c'_2}{c'_1} Y_{3,1}^{(0)} & -\frac{c'_2}{c'_1} Y_{3,3}^{(0)} & -\frac{c'_2}{c'_1} Y_{3,2}^{(0)} \\
        -\frac{c'_2}{c'_1} Y_{3,3}^{(0)} & 2 \frac{c'_2}{c'_1} Y_{3,2}^{(0)} & 1-\frac{c'_2}{c'_1} Y_{3,1}^{(0)} \\
        -\frac{c'_2}{c'_1} Y_{3,2}^{(0)} & 1-\frac{c'_2}{c'_1} Y_{3,1}^{(0)} & 2 \frac{c'_2}{c'_1} Y_{3,3}^{(0)}
    \end{pmatrix},\quad M_D = c_1 \upsilon \begin{pmatrix}
        Y_1^{(-2)} + 2 \frac{c_2}{c_1} Y_{3,1}^{(-2)} & 0 & -2 \frac{c_2}{c_1} Y_{3,2}^{(-2)} \\
        -2 \frac{c_2}{c_1} Y_{3,3}^{(-2)} & 2 \frac{c_2}{c_1} Y_{3,2}^{(-2)} &  Y_1^{(-2)} \\
        0 & Y_1^{(-2)} - 2 \frac{c_2}{c_1} Y_{3,1}^{(-2)} & 2 \frac{c_2}{c_1} Y_{3,3}^{(-2)}
    \end{pmatrix}
\end{equation}
From the type I seesaw formula, $m_\nu = -M_D^T M_R^{-1} M_D$, we obtain an additional overall mass scale given by $(c_1 \upsilon)^2 / c'_1 \Lambda$. As a result, model I contains a total of 21 independent real parameters, including both the real and imaginary parts of the modulus $\tau$.

\medskip
\textbf{Model II:}
In this model, the $A_4$ charge assignments for the matter fields in the $10_{F_i}$ and $\bar{5}{F_i}$ representations remain unchanged from Model I. However, the RH neutrinos have different assignments, transforming as $A_4$ singlets with charges $(N_1^c, N_2^c, N_3^c) \sim (1, 1'', 1')$. The weights for $T_1, T_2$, and $T_3$ are the same as in Model I, leading to a similar up-quark mass matrix. In contrast, the different weight of $\bar{5}{F_i}$ results in modified mass matrices for charged leptons, down quarks, and neutrino Dirac masses. See Table \ref{tab2} for the $A_4$ charge and weight assignments of all fields, along with the polyharmonic Maa{\ss} forms relevant for this model.
\begin{table}[H]
\centering
\setlength{\tabcolsep}{8pt}
\begin{tabular}{|c||c|c|c|c|c|c|c|c|c|c|c|c|c|c|}
\hline
Model II & $T_1$ & $T_2$ & $T_3$ & $\bar{F}_i$ & $N_1^c$ & $N_2^c$ & $N_3^c$ & $H_5^\ast$ & $H_{45}^\ast$ & $Y_{\mathbf{3}}^{(-4)}$ & $Y_{\mathbf{3}}^{(-2)}$ & $Y_{\mathbf{1}}^{(4)}$ & $Y_{\mathbf{1'}}^{(4)}$ & $Y_{\mathbf{1}}^{(6)}$ \\ \hline
$SU(5)$ & $10$ & $10$ & $10$ & $\bar{5}$ & $1$ & $1$ & $1$ & $\bar{5}$ & $\bar{45}$ & $1$ & $1$ & $1$ & $1$ & $1$ \\ \hline
$A_4$ & $\mathbf{1''}$ & $\mathbf{1'}$ & $\mathbf{1}$ & $\mathbf{3}$ & $\mathbf{1}$ & $\mathbf{1''}$ & $\mathbf{1'}$ & $\mathbf{1}$ & $\mathbf{1}$ & $\mathbf{3}$ & $\mathbf{3}$ & $\mathbf{1}$ & $\mathbf{1'}$ & $\mathbf{1}$ \\ \hline
$k_I$ & $4$ & $2$ & $0$ & $-4$ & $0$ & $2$ & $2$ & $0$ & $0$ & $-4$ & $-2$ & $4$ & $4$ & $6$ \\ \hline
\end{tabular}
\caption{Same as Table \ref{tab1} but for model II.}
\label{tab2}
\end{table}
Given the requirement of invariance under both the $SU(5)$ and the modular symmetry $\Gamma_3$, the renormalizable Yukawa Lagrangian takes the form
\begin{eqnarray}
    -{\cal{L}}_Y^{II} &=& a_1 (T_1 T_2)_{\mathbf{1}} Y_\mathbf{1}^{(6)} H_5 + a_2 (T_1 T_3)_{\mathbf{1''}} Y_\mathbf{1'}^{(4)} H_5 + a_3 (T_2 T_2)_{\mathbf{1''}} Y_\mathbf{1'}^{(4)} H_5 + a_4 (T_3 T_3)_{\mathbf{1}} H_5 + b_1 T_1 (\bar{F}_j Y_\mathbf{3}^{(0)})_\mathbf{1'} H_5^\ast  \nonumber \\
    &+& b_2 T_2 (\bar{F}_j Y_\mathbf{3}^{(-2)})_\mathbf{1''} H_5^\ast + b_3 T_3 (\bar{F}_j Y_\mathbf{3}^{(-4)})_\mathbf{1} H_5^\ast + b'_1 T_1 (\bar{F}_j Y_\mathbf{3}^{(0)})_\mathbf{1'} H_{45}^\ast + b'_2 T_2 (\bar{F}_j Y_\mathbf{3}^{(-2)})_\mathbf{1''} H_{45}^\ast \nonumber \\
    &+& b'_3 T_3 (\bar{F}_j Y_\mathbf{3}^{(-4)})_\mathbf{1} H_{45}^\ast + c_1 (\bar{F}_i N_1^c)_\mathbf{3} Y_\mathbf{3}^{(-4)} H_5 + c_2 (\bar{F}_i N_2^c)_\mathbf{3} Y_\mathbf{3}^{(-2)} H_5 + c_3 (\bar{F}_i N_3^c)_\mathbf{3} Y_\mathbf{3}^{(-2)} H_5  \\
    &+& c'_1 \Lambda_1 (N_1^c N_1^c)_\mathbf{1} + c'_2 \Lambda_2 (N_2^c N_3^c)_\mathbf{3} Y_\mathbf{1}^{(4)} + c'_3 \Lambda_2 (N_3^c N_3^c)_\mathbf{3} Y_\mathbf{1'}^{(4)} + h.c. \nonumber
    \label{ly}
\end{eqnarray}
Using the decomposition of the tensor product of $A_4$ representations from Eq. \ref{dypr}, we derive the Majorana mass matrix and the Yukawa matrices for this model We find
\begin{eqnarray}
    Y_{\bar{5}} = \begin{pmatrix}
        b_1 Y_{3,2}^{(0)} & b_2 Y_{3,3}^{(-2)} & b_3 Y_{3,1}^{(-4)} \\ b_1 Y_{3,1}^{(0)} & b_2 Y_{3,2}^{(-2)} & b_3 Y_{3,3}^{(-4)} \\
        b_1 Y_{3,3}^{(0)} & b_2 Y_{3,1}^{(-2)} & b_3 Y_{3,2}^{(-4)}
    \end{pmatrix}, \; Y_D = \begin{pmatrix}
        c_1 Y_{3,1}^{(-4)} & c_1 Y_{3,3}^{(-4)} & c_1 Y_{3,2}^{(-4)} \\
        c_2 Y_{3,2}^{(-2)} & c_2 Y_{3,1}^{(-2)} & c_2  Y_{3,3}^{(-2)} \\ c_3 Y_{3,3}^{(-2)} & c_3 Y_{3,2}^{(-2)} & c_3 Y_{3,1}^{(-2)}
    \end{pmatrix}, \; M_R = c'_1 \Lambda \begin{pmatrix}
        1 & 0 & 0 \\
        0 & 0 & \frac{c'_2}{c'_1} Y_1^{(4)} \\
        0 & \frac{c'_2}{c'_1} Y_1^{(4)} & \frac{c'_3}{c'_1} Y_{1'}^{(4)}
    \end{pmatrix}
    \label{ym2}
\end{eqnarray}
The Yukawa matrix for the up-type quarks, $Y_{10}$, is the same as the one derived in model I (see Eq. \ref{ycou}). On the other hand, the Yukawa matrix associated with the 45-dimensional Higgs, $Y_{\bar{45}}$, closely resembles $Y_{\bar{5}}$, except that the couplings $b_i$ are replaced by $b'_i$. As in model I, the couplings $a_3,~b_1,~b_2,~b_3,~b'_1,~c_1,~c_2,~c_3,~c'_1$ are unphysical, while the remaining couplings are complex. Consequently, the model contains a total of 24 free parameters.
In the following section, we carry out a detailed numerical analysis of the two proposed models, identifying regions in the parameter space that show excellent agreement with experimental data for both the lepton and quark sectors.
\section{Numerical results}
\label{sec4}
In this section, we provide a detailed numerical analysis of fermion flavor structure predictions for the two benchmark $SU(5) \otimes \Gamma_3$ models introduced earlier. These models differ in their modular weight assignments and the $A_4$ transformation properties of certain matter fields as described in Tables \ref{tab1} and \ref{tab2}. Our study employs polyharmonic Maa{\ss} forms of level 3, incorporating multiplets with modular weights $k_Y = -4, -2, 0, 2, 4, 6$. The Yukawa matrices, given in Eqs. \ref{ycou} and \ref{ym2}, depend on modulus $\tau$, coupling constant ratios, and mass scales, which govern fermion masses, mixings, and {\it CP} violating phases. Our analysis is conducted at the GUT scale, using quark and charged lepton masses and CKM mixing angles from Ref. \cite{Babu:2016bmy}. For neutrino oscillation parameters, we adopt the latest NuFIT v6.0 global fit, which includes SK atmospheric data \cite{Esteban:2024eli}. Therefore, their RGE effects from the low-energy to the GUT scale are not considered. Numerical values for all the parameters can be found in Table \ref{data}.
\begin{table}[H]
\centering
\renewcommand{\arraystretch}{1.5}
\setlength{\tabcolsep}{10pt}
\begin{tabular}{|c|c||c|c|c|}
\hline
Parameters & $\mu_i \pm 1\sigma$ & Parameters & $\mu_i \pm 1\sigma$ & $3\sigma$ ranges \\ \hline
$m_e/m_\mu$ & $0.004737 \pm 0.000007$  & $\Delta m_{21}^2/10^{-5}eV^2$ & $7.49 \pm 0.19$ & $6.92 \rightarrow 8.05$ \\ \hline
$m_\mu/m_\tau$ & $0.058823 \pm 0.000083$ & $\Delta m_{31}^2/10^{-3}eV^2$(NO) & $2.513_{-0.019}^{+0.021}$ & $2.451 \rightarrow 2.578$ \\ \hline
$m_u/m_c$ & $0.001857 \pm 0.000629$ & $\Delta m_{32}^2/10^{-3}eV^2$(IO) & $-2.484 \pm 0.020$ & $-2.547 \rightarrow -2.421$ \\ \hline
$m_c/m_t$ & $0.003194 \pm 0.000098$ & $\sin^2\theta_{12}^l$ & $0.308_{-0.011}^{+0.012}$ & $0.275 \rightarrow 0.345$ \\ \hline
$m_d/m_s$ & $0.052827 \pm 0.005812$ & $\sin^2\theta_{23}^l$(NO) & $0.470_{-0.013}^{+0.017}$ & $0.435 \rightarrow 0.585 $ \\ \hline
$m_s/m_b$ & $0.021710 \pm 0.001163$ & $\sin^2\theta_{23}^l$(IO) & $0.550_{-0.015}^{+0.012}$ & $0.440 \rightarrow 0.584$ \\ \hline
$\theta_{12}^q$ & $0.2274 \pm 0.00052$ & $\sin^2\theta_{13}^l$(NO) & $0.02215_{-0.00058}^{+0.00056}$ & $0.02030 \rightarrow 0.02388$ \\ \hline
$\theta_{13}^q$ & $0.0042 \pm 0.00017$ & $\sin^2\theta_{13}^l$(IO) & $0.02231 \pm 0.00056$ & $0.02060 \rightarrow 0.02409$ \\ \hline
$\theta_{23}^q$ & $0.0485 \pm 0.00070$ & $\delta_{CP}^l/^\circ$(NO) & $212_{-41}^{+26}$ & $124 \rightarrow 364$ \\ \hline
$\delta_{CP}^q/^\circ$ & $69.16 \pm 7.89$ & $\delta_{CP}^l/^\circ$(IO) & $274_{-25}^{+22}$  & $201 \rightarrow 335$ \\ \hline
\end{tabular}
\caption{Best-fit values and $1\sigma$ uncertainties for charged fermion mass ratios, quark mixing angles, and the Dirac {\it CP}-violating phase at the GUT scale ($M_{GUT} = 2\times 10^{16}~\text{GeV}$), taken from Ref. \cite{Babu:2016bmy}. The lepton mixing angles, neutrino mass squared differences, and leptonic {\it CP}-violating phase from the latest global fit NuFIT v6.0, including SK atmospheric data for normal and inverted neutrino mass ordering \cite{Esteban:2024eli}.}
\label{data}
\end{table}
The standard parametrization is adopted for both the lepton and quark mixing matrices. The unitary matrix $U$ follows the PDG convention and is given by \cite{PDG}  
\begin{equation}  
U =  
\begin{pmatrix}  
c_{12} c_{13} & s_{12} c_{13} & s_{13} e^{-i \delta_{CP}} \\  
-s_{12} c_{23} - c_{12} s_{13} s_{23} e^{i \delta_{CP}} & c_{12} c_{23} - s_{12} s_{13} s_{23} e^{i \delta_{CP}} & c_{13} s_{23} \\  
s_{12} s_{23} - c_{12} s_{13} c_{23} e^{i \delta_{CP}} & -c_{12} s_{23} - s_{12} s_{13} c_{23} e^{i \delta_{CP}} & c_{13} c_{23}  
\end{pmatrix}.  
\end{equation}
For the PMNS matrix, $\theta_{ij} \equiv \theta_{ij}^{l}$ and $\delta_{CP} \equiv \delta_{CP}^{l}$, with an additional diagonal phase matrix $\text{diag}(1, e^{i\alpha_{21}/2}, e^{i\alpha_{31}/2})$ accounting for the Majorana phases. For the CKM matrix, $\theta_{ij} \equiv \theta_{ij}^{q}$ and $\delta_{CP} \equiv \delta_{CP}^{q}$. We analyze the predictions of the two benchmark $SU(5) \times \Gamma_3$ models for charged fermion mass ratios, quark-sector mixing, {\it CP} violating phases, and neutrino oscillation parameters. The Majorana phases in the PMNS matrix directly impact the phenomenology of neutrino mass constraints. The absolute neutrino mass scale can be probed using three different observables:
\medskip

\textbf{(i)} Sum of three neutrino masses $\sum m_i$ from cosmological observations: Within the $\Lambda$CDM framework, the latest Planck cosmic microwave background (CMB) measurements, which include the Planck power spectra and Planck lensing (TT, TE, EE + lowE + lensing), set an upper bound on the sum of neutrino masses at $\sum m_i < 0.24~\text{eV}$ (95\% CL) \cite{Planck:2018vyg}, which tightens to $\sum m_i < 0.12~\text{eV}$ when combined with baryon acoustic oscillation (BAO) data \cite{Planck:2018vyg}. These bounds can be translated into limits on the lightest neutrino mass ($m_1$ for normal ordering, $m_3$ for inverted ordering) by expressing the total mass in terms of the mass-squared differences 
\begin{eqnarray} \sum m_i &=& m_1 + \sqrt{m_1^2 + \Delta m_{21}^2} + \sqrt{m_1^2 + \Delta m_{31}^2} \quad \text{(NO)}, \nonumber \\ \sum m_i &=& m_3 + \sqrt{m_3^2 + |\Delta m_{32}^2|} + \sqrt{m_3^2 + |\Delta m_{32}^2| - \Delta m_{21}^2} \quad \text{(IO)}.
\label{csn}
\end{eqnarray} 
Using the best fit values of $\Delta m_{ij}^2$ from Table~\ref{data}, we obtain $m_1 < 26~\text{meV}$ (NO) and $m_3 < 15.6~\text{meV}$ (IO) from Planck CMB data alone and $m_1 < 79.8~\text{meV}$ (NO) and $m_3 < 77.6~\text{meV}$ when combining Planck CMB with BAO data.

\textbf{(ii)} The effective Majorana mass $m_{\beta\beta}$ in neutrinoless double beta decay experiments: This process is a key probe of the Majorana nature of neutrinos. Its decay rate depends on the Majorana {\it CP} phases and the effective mass $|m_{\beta\beta}|$ defined as $|m_{\beta\beta}| = \left|\sum_i U_{e_i}^2 m_i \right|$. KamLAND-Zen experiment sets the most stringent limit on $0\nu\beta\beta$-decay half-life, which provides an upper bound on $|m_{\beta\beta}|$ given by $|m_{\beta\beta}| < (28 \sim 122)~\text{meV}$ \cite{KamLAND-Zen:2024eml}. Future large-scale $0\nu\beta\beta$-decay experiments, such as LEGEND-1000 and nEXO, aim to improve sensitivity, with LEGEND-1000 targeting $m_{\beta\beta}$ in the range of $9-21~\text{meV}$ \cite{LEGEND:2021bnm}, and nEXO aiming for $4.7-20.3~\text{meV}$ \cite{nEXO:2021ujk}.

\textbf{(iii)} The effective neutrino mass $m_\beta$ in beta decay experiments: Tritium beta decay provides a direct probe of the neutrino mass scale by analyzing the electron energy spectrum near its endpoint. The effective anti-electron neutrino mass squared is defined as $m_\beta^2 = \sum_{i=1}^3 |U_{e_i}|^2 m_i^2$. The KATRIN collaboration currently constrains $m_\beta < 0.45~\text{eV}$ \cite{Katrin:2024tvg}, with a future sensitivity goal of $0.2~\text{eV}$ \cite{KATRIN:2021dfa}.
\begin{table}
\begin{tabular}{c||c|c|c|c}
Parameters & Model I (NO) & Model I (IO) & Model II (NO) & Model II (IO)  \\
\hline
$Re(\tau)$ & $0.1281$ & $0.0103$ & $0.0476$  & $-0.2831$   \\
$Im(\tau)$ & $1.6823$ & $2.2648$ & $1.7104$  & $1.6590$   \\
$a_1/a_3$ & $0.1443 e^{0.2925 \pi i}$ & $0.46934 e^{-0.50701 \pi i}$ & $1.1834e^{0.9584\pi i}$ & $1.3868 e^{0.8950 \pi i}$   \\
$a_2/a_3$ & $7.681 e^{0.6975 \pi i}$ & $357.094 e^{-0.37838 \pi i}$ & $228.6683e^{-0.2669\pi i}$ & $249.2469 e^{-0.0387 \pi i}$   
\\
$a_3 \upsilon~(\text{GeV})$ & $38.5662$ & $68.2266$ & $165.9036$ & $68.2602$ 
\\
$a_4/a_3$ & $128.90 e^{-0.9322 \pi i}$ & $675.274 e^{0.28897 \pi i}$ & $1398.5939e^{-0.3610\pi i}$ & $1651.0574 e^{-0.2782 \pi i}$ 
\\
$b_1/b_3$ & $1.5381$ & $1.3174$ & $0.2582$ & $0.0521$ 
\\
$b_2/b_3$ & $100.5243$ & $-984.3330$ & $-87.9918$ & $-10.5364$ 
\\
$b_3 \upsilon~(\text{GeV})$ & $57.8613$ & $0.0973$ & $86.7449$ & $166.0648$
\\
$b'_1/b_3$ & $0.5024$ & $0.3122$ & $0.1273$ & $0.0343$ 
\\
$b'_2/b_3$ & $6.874 e^{0.9384 \pi i}$ & $97.3373 e^{-0.61665 \pi i}$ & $97.9927e^{-0.7851\pi i}$ & $35.7346 e^{-0.8508 \pi i}$ 
\\
$b'_3/b_3$ & $0.9314 e^{0.9985 \pi i}$ & $6.9251 e^{-0.30634 \pi i}$ & $2.5877e^{0.6506\pi i}$ & $0.9658 e^{0.7905 \pi i}$  
\\
$c_2/c_1$ & $0.1516$ & $-9.1$ & $722.5162$ & $185.3394$ 
\\
$c_3/c_1$ & $---$ & $---$ & $-883.9098$ & $-1014.7310$ 
\\
$c'_2/c'_1$ & $2.458 e^{0.2027 \pi i}$ & $0.94975 e^{0.00121 \pi i}$ & $185442.2760e^{0.3944\pi i}$ & $105888.2838 e^{0.4987 \pi i}$ 
\\
$c'_3/c_1$ & $---$ & $---$ & $216055.8955e^{-0.0428\pi i}$ & $940437.4129 e^{-0.0432 \pi i}$ 
\\
$\frac{(c_1 \upsilon)^2}{c'_1 \Lambda}~(\text{meV})$ & $124.1089$  & $0.0398$ & $4.8530$ & $9.1278$ 
\\
\hline
$m_e/m_\mu$ & $0.004737$ & $0.004747$ & $0.004736$ & $0.004746$ 
\\
$m_\mu/m_\tau$ & $0.0588$ & $0.0588$ & $0.0588$ & $0.0588$
\\
$m_1~(\text{meV})$ & $76.00$ & $50.87$ & $40.55$ & $57.55$ 
\\
$m_2~(\text{meV})$ & $76.49$ & $51.60$ & $41.46$ & $58.20$ 
\\
$m_3~(\text{meV})$ & $91.02$ & $13.40$ & $64.47$ & $30.09$ 
\\
$\sin^2\theta^l_{12}$ & $0.308$ & $0.308$ & $0.308$ & $0.308$
\\
$\sin^2\theta^l_{13}$ & $0.02249$ & $0.02281$ & $0.02217$ & $0.02217$ 
\\
$\sin^2\theta^l_{23}$ & $0.5109$ & $0.7838$ & $0.4536$ & $0.6096$ 
\\
$\delta_{CP}^l/\pi$ & $1.599$ & $1.475$ & $0.522$ & $1.467$ 
\\
$\alpha_{21}/\pi$ & $1.544$ & $1.582$ & $1.506$ & $1.622$ 
\\
$\alpha_{31}/\pi$ & $1.574$ & $0.428$ & $0.433$ & $1.576$ 
\\
$m_\beta~(\text{meV})$ & $76.53$ & $51.54$ & $41.52$ & $58.16$ 
\\
$m_{\beta\beta}~(\text{meV})$ & $131.28$ & $45.35$ & $40.479$ & $56.56$ 
\\
\hline
$m_u/m_c$ & $0.001860$ & $0.001874$ & $0.001856$ & $0.001886$ 
\\
$m_c/m_t$ & $0.003191$ & $0.003192$ & $0.003193$ & $0.003353$ 
\\
$m_d/m_s$ & $0.053783$ & $0.054773$ & $0.052822$ & $0.04743$ 
\\
$m_s/m_b$ & $0.02306$ & $0.020313$ & $0.02171$ & $0.02158$ 
\\
$\theta^q_{12}$ & $0.2273$ & $0.2272$ & $0.2274$ & $0.2272$ 
\\
$\theta^q_{13}$ & $0.0041$ & $0.0042$ & $0.0042$ & $0.0042$
\\
$\theta^q_{23}$ & $0.04877$ & $0.04757$ & $0.04851$ & $0.04814$ 
\\
$\delta_{CP}^q/^\circ$ & $69.08$ & $60.93$ & $69.14$ & $61.60$ 
\\
\hline
$\chi_l^{2}$ & $0.737883$ & $11.384881$ & $1.013631$ & $0.715045$ 
\\
$\chi_q^{2}$ & $0.172793$ & $0.069399$ & $0.000014$ & $1.115284$ 
\\
$\chi_{totlal}^{2}$ & $0.910676$ & $11.888987$ & $1.013645$ & $1.830329$  
\end{tabular}
\caption{The best-fit values of the model parameters, along with the corresponding predictions for fermion masses and mixing, are presented for the two benchmark $SU(5) \times \Gamma_3$ models in both NO and IO. The numerical values of the physical observables are summarized in Table \ref{data}.}
\label{results}
\end{table}

To assess the phenomenological viability of our benchmark models for both NO and IO, we perform a standard $\chi^2$ analysis, with the $\chi^2$ function defined as  
\begin{equation}  
\chi^2 = \sum_i \left( \frac{Q_i(p) - \mu_i}{\sigma_i} \right)^2,  
\end{equation}  
where $\mu_i$ and $\sigma_i$ denote the global best-fit values and corresponding $1\sigma$ uncertainties, respectively, for the 16 flavor observables listed in Table \ref{data}. The quantities $Q_i(p)$ represent the theoretical predictions for these observables, expressed as functions of the free parameters of each model $p = \{\tau, a_i/a_3, b_i/b_3, b'_i/b_3, \dots \}$. The modulus $\tau$ is considered as a free parameter where its value varies within the fundamental domain ${\cal{D}} = \{\tau \in \mathbb{C} | Im(\tau)>0, |Re(\tau)| \leq \frac{1}{2}, |\tau| \geq 1 \}$. The total $\chi^2$ can be thought of as the sum of two contributions: $\chi^2_l$ for leptonic observables and $\chi^2_q$ for quark observables. Specifically, $\chi^2_l$ is constructed from charged lepton mass ratios and neutrino oscillation parameters, while $\chi^2_q$ is based on quark mass ratios, quark mixing angles, and the Dirac {\it CP} violating phase as listed in Table \ref{data}.
\begin{figure}[H]
    \centering
    \includegraphics[width=1\textwidth]{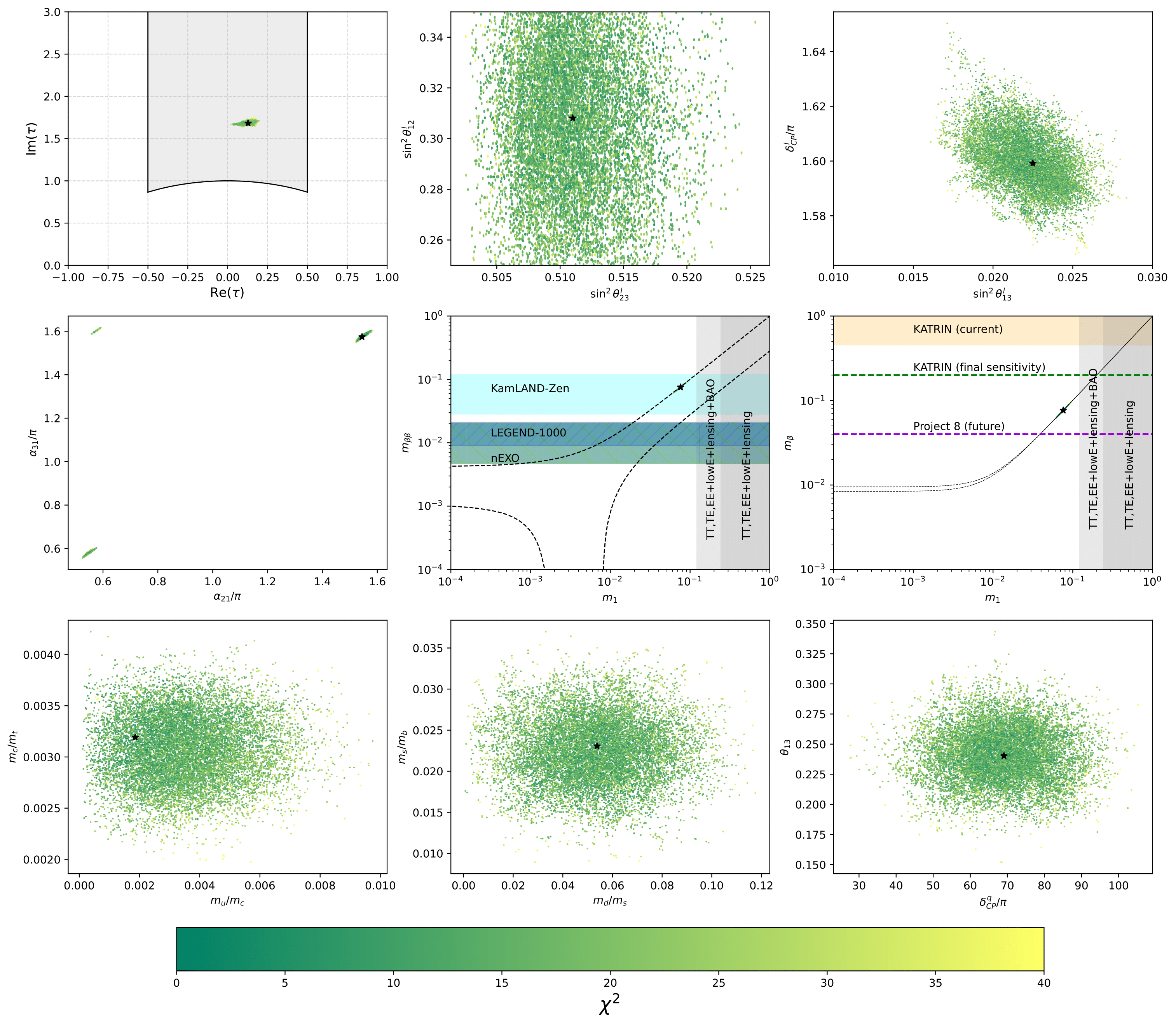}
    \caption{Allowed parameter space for $\tau$ and predicted correlations among quark and lepton observables in model I for the NO case. The black star indicates the best-fit point. In the second row, the horizontal shaded regions in the middle and right panels show experimental constraints on the sum of neutrino masses. The violet region represents the latest Planck CMB limit, $\sum m_i < 0.24~\text{eV}$ at 95\% CL \cite{Planck:2018vyg}, while the gray region shows the tighter limit, $\sum m_i < 0.12~\text{eV}$, obtained by combining Planck and BAO data \cite{Planck:2018vyg}. The gray and violet dashed lines translate these limits into bounds on the lightest neutrino mass, obtained using Eq.~\ref{csn}. Vertical shaded regions and lines indicate experimental constraints from beta decay and neutrinoless double-beta ($0\nu\beta\beta$) decay experiments.}
    \label{fig1}
\end{figure}
To determine the masses and mixing angles of quarks and leptons, we apply singular value decomposition to the mass matrices and then compute the total $\chi^2_{\text{total}}$. All dimensionless parameters are treated as independent variables, with their magnitudes randomly varied within $[0, 10^4]$, while their phases are varied in the range $[0, 2\pi]$. We numerically minimize the $\chi^2_{total}$ function using the \texttt{FlavorPy} package~\cite{FlavorPy} to determine the parameter values that best align with experimental data, thereby evaluating the validity of each model\footnote{A similar procedure was recently employed in Ref. \cite{Belfkir:2024uvj} within the context of a SUSY modular flavor model.}. The package utilizes the \texttt{lmfit} algorithm for parameter optimization and conducts a Markov Chain Monte Carlo (MCMC) scan around the best-fit point to explore parameter uncertainties. A model is considered phenomenologically viable if it predicts all or nearly all of the 16 observables listed in Table \ref{data} within their $3\sigma$ experimental limits. This requires a sufficiently low $\chi^2_{total}$, and we restrict our analysis to models satisfying $\chi^2_{total} < 10$. 
\begin{figure}[H]
    \centering
    \includegraphics[width=1\textwidth]{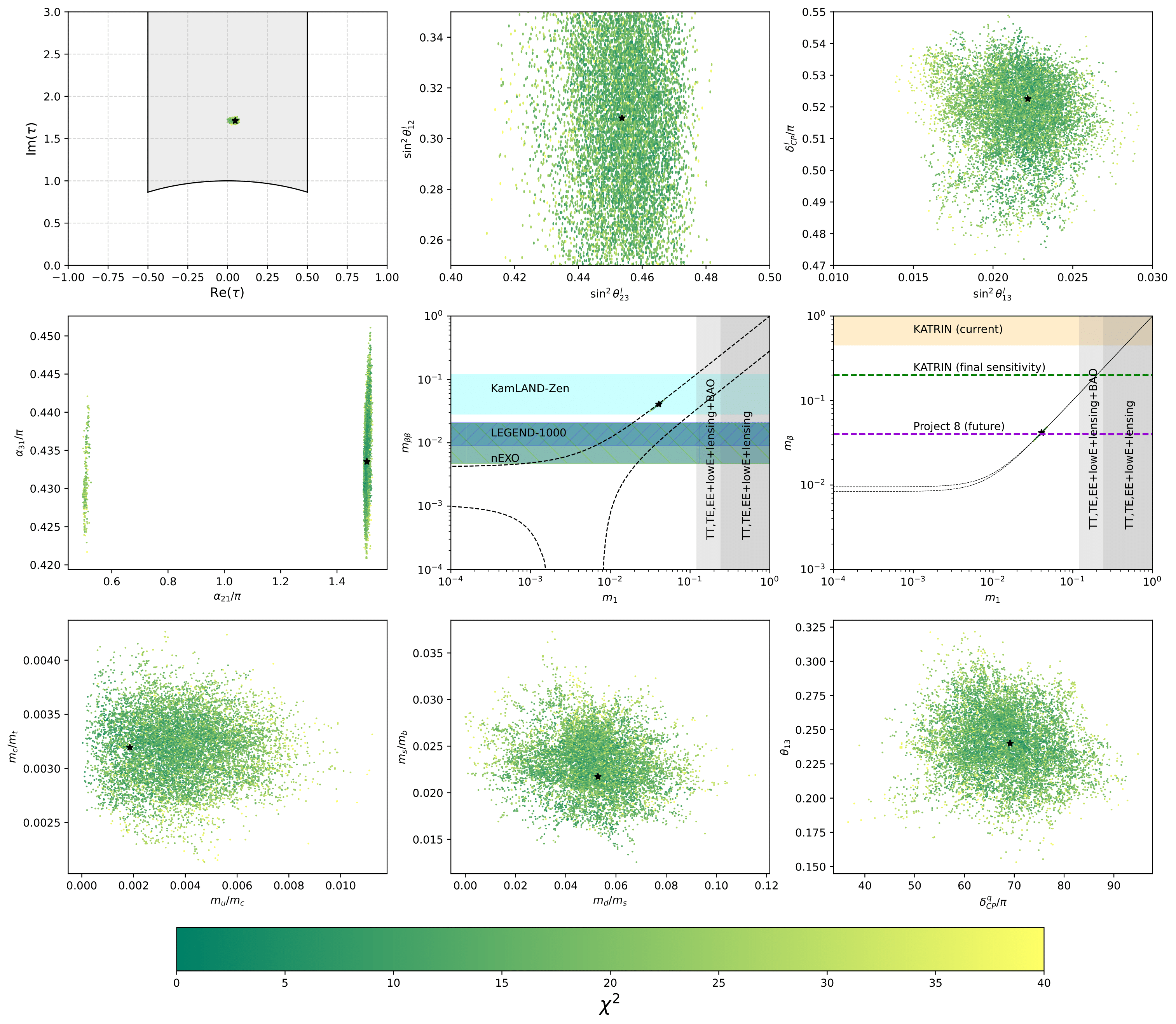}
    \caption{Same as in figure \ref{fig1} but for model II.}
    \label{fig2}
\end{figure}
Beyond the benchmark models I and II discussed in the previous section, we investigated alternative configurations with different $A_4$ charge assignments for the matter fields. However, these models failed to yield viable solutions. For instance, when the 10-dimensional representation was assigned as an $A_4$ triplet and the three generations of the five-dimensional representation $\bar{5}_{F_i}$ were treated as $A_4$ singlets--while keeping the scalar sector trivial--no viable solutions emerged. Regardless of the weight assignments, which modify the resulting mass matrices, all cases resulted in $\chi^2_{total}$ values exceeding 50, with several observables falling outside their experimentally allowed $3\sigma$ range. The best-fit values of the free parameters for benchmark models I and II are summarized in Table \ref{results} for both neutrino mass hierarchies. The table also includes predictions for key observables, such as fermion mass ratios, flavor mixing parameters, the effective Majorana neutrino mass $m_{\beta\beta}$, the effective electron antineutrino mass $m_\beta$, as well as the three light neutrino masses $m_{i=1,2,3}$. Additionally, the table provides the minimum values of the $\chi^2$ functions $\chi_l^2$, $\chi_q^2$ and $\chi^2_{total}$ for both mass ordering. As shown in the last row of Table \ref{results}, the NO neutrino mass spectrum yields the lowest $\chi^2_{total}$ values for both models. In contrast, the IO spectrum results in a high $\chi^2_{total}>10$ for model I but remains within $3\sigma$ agreement with data for model II with $\chi^2_{total} = 1.8303$. This suggests that model I favors only normal ordering, while model II allows both mass orderings.
\begin{figure}[H]
    \centering
    \includegraphics[width=1\textwidth]{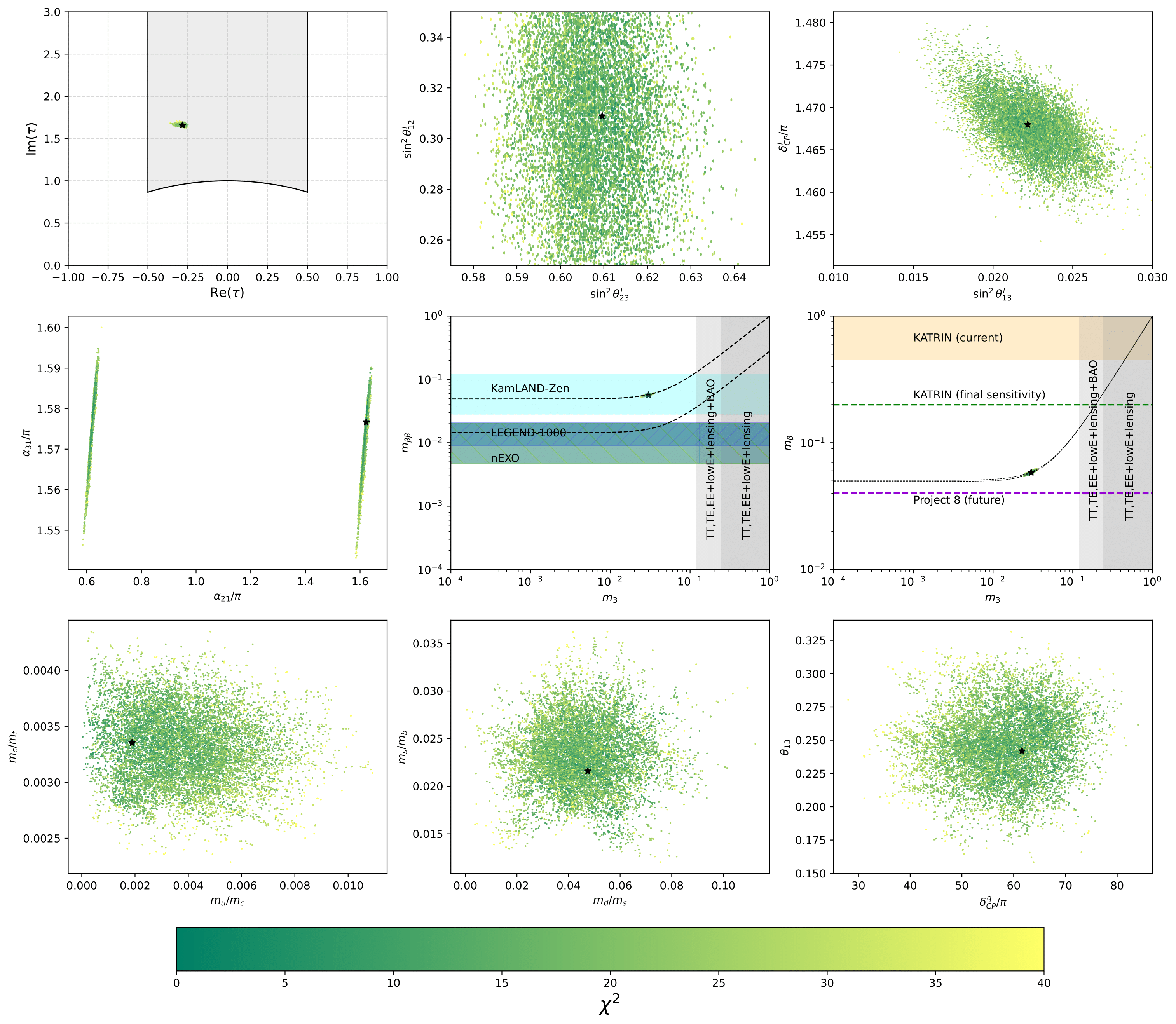}
    \caption{Same as in figure \ref{fig2} but for the inverted neutrino mass ordering.}
    \label{fig3}
\end{figure}
Model I with a normal mass ordering is consistent with the experimental data at the $3\sigma$ level, resulting in a total chi-square value of $\chi^2_{\text{total}} = 0.9106$. Among the 16 fitted observables, 12 fall within their $1\sigma$ experimentally allowed ranges, while the remaining four are within their $3\sigma$ limits. For model II with a normal mass ordering, we obtain $\chi^2_{\text{total}} = 1.0136$, where 14 observables lie within their $1\sigma$ ranges. The observable $\sin^2\theta_{23}^l$ is within its $3\sigma$ range, while $\delta_{CP}^l$, the least constrained oscillation parameter, falls outside the $3\sigma$ bound.  In the case of model II with an inverted mass ordering, the fit yields $\chi^2_{\text{total}} = 1.8303$. Here, 13 observables are within their $1\sigma$ ranges, while two mass ratio observables ($m_e/m_\mu$ and $m_c/m_t$) are within their $3\sigma$ limits. However, $\sin^2\theta_{23}^l$ lies outside the $3\sigma$ range. The correlations between the observables used in our numerical analysis are depicted in Figures \ref{fig1}, \ref{fig2}, and \ref{fig3}, corresponding to models I, II with NO, and model II with inverted ordering, respectively.
In each figure, the first plot displays the allowed values of the modulus $\tau$ within the fundamental domain $\mathcal{D}$, represented by the gray region, with the best fit value marked by a black star. For the NO case, $\tau$ is confined to a narrow region on the right side of $\mathcal{D}$ in both models I and II (Figs. \ref{fig1} and \ref{fig2}), while for the IO case in model II (\ref{fig3}), it is located on the left side of $\mathcal{D}$. 
The second plot in the first row illustrates the correlation between $\sin^2\theta^l_{12}$ and $\sin^2\theta^l_{23}$. In model I (Figure \ref{fig1}), the atmospheric angle lies in the higher octant, with the best fit value at $\sin^2\theta^l_{23} = 0.510$. In model II with NO, $\sin^2\theta^l_{23}$ is primarily in the lower octant, with a best fit value of $\sin^2\theta^l_{23} = 0.453$, whereas for the IO case, it lies in the higher octant but falls outside the $3\sigma$ range from NuFit 6.0. The best-fit value of the solar angle in all models remains consistent with the NuFit 6.0 result: $\sin^2\theta^l_{12} = 0.308$. The third figure in the first row presents the correlation between $\delta_{CP}^l$ and $\sin^2\theta_{13}$. Model II predicts the leptonic Dirac CP phase most accurately in the IO case but falls outside the $3\sigma$ range in the NO case, whereas Model I remains within this $3\sigma$ range. The best-fit value of the reactor angle is consistently within $1\sigma$ level for all models.

The second row displays three plots showing the correlation between the two Majorana {\it CP} phases, $\alpha_{21}$ and $\alpha_{31}$, and the effective neutrino masses, $m_{\beta\beta}$ and $m_\beta$, as functions of the lightest neutrino mass; $m_1$ for NO and $m_3$ for IO. The Majorana phases are well constrained in this study and deviate significantly from the {\it CP}-conserving values of $0$ and $\pi$. In all models, the best-fit values are mostly around $\pi/2$. Specifically, in model I with NO, the phases are $\alpha_{21} \approx 1.544\pi$ and $\alpha_{31} \approx 1.574\pi$. In model II with NO, the values are $\alpha_{21} \approx 1.506\pi$ and $\alpha_{31} \approx 0.433\pi$. For model II with IO, the phases are $\alpha_{21} \approx 1.622\pi$ and $\alpha_{31} \approx 1.576\pi$. For the effective neutrino masses $m_{\beta\beta}$ and $m_\beta$, the plots for all models indicate that these parameters are tightly constrained, as the data points are closely clustered around their best-fit values. Below, we present the best-fit values for $m_{\beta\beta}$, $m_\beta$, the lightest neutrino mass, and the total sum of neutrino masses for each model
\begin{eqnarray}
    \label{mbbb}
    &\text{Model I (NO):}& \quad m_{\beta\beta} = 131.28~\text{meV}, \mkern9mu m_\beta = 76.53~\text{meV}, \mkern9mu m_1 = 76.00~\text{meV}, \mkern9mu \sum m_i = 243.52~\text{meV} \nonumber \\
    &\text{Model II (NO):}& \quad m_{\beta\beta} = 40.47~\text{meV}, \mkern9mu m_\beta = 41.52~\text{meV}, \mkern9mu m_1 = 40.55~\text{meV}, \mkern9mu \sum m_i = 146.49~\text{meV} \\
    &\text{Model II (IO):}& \quad m_{\beta\beta} = 56.56~\text{meV}, \mkern9mu m_\beta = 58.16~\text{meV}, \mkern9mu m_3 = 30.09~\text{meV}, \mkern9mu \sum m_i = 145.84~\text{meV}. \nonumber
\end{eqnarray}
From the plots, it is evident that for all models, the predicted values of $m_{\beta\beta}$ fall within the region bounded by dashed lines, which correspond to the allowed range of $m_{\beta\beta}$ derived from the $3\sigma$ uncertainties in neutrino oscillation parameters. Moreover, these predictions remain consistent with the current most stringent constraint from the KamLAND-Zen collaboration represented by the cyan-shaded region. For $m_\beta$, the best-fit values across all models remain below both the current (orange shaded region) and future (green dashed line) projected sensitivities of the KATRIN experiment. If KATRIN fails to detect $m_\beta$ within its future sensitivity limits, indicating that $m_\beta < 0.2~\text{eV}$, our predicted best-fit values could be tested by the upcoming Project 8 experiment, which aims to achieve a significantly improved sensitivity of approximately $0.04~\text{eV}$ \cite{Project8:2017nal}. In model I (NO), the best-fit value of the sum of neutrino masses in Eq.\ref{mbbb} is close to the upper limit set by the latest Planck CMB constraints, $\sum m_i < 0.24\text{eV}$. The corresponding lightest neutrino mass, $m_1 = 76~\text{meV}$, is consistent with the upper bound $m_1 < 79.8~\text{meV}$ derived from translating the Planck constraint, as discussed following Eq.\ref{csn}. In model II, for both normal and inverted hierarchies, the sum of neutrino masses remains well below this Planck limit, satisfying the constraint. Additionally, the lightest neutrino masses also lie below the upper bounds derived from the $\sum m_i < 0.24\text{eV}$, with $m_1 < 79.8\text{meV}$ for NO and $m_3 < 77.6~\text{meV}$ for IO.
However, all models predict a best-fit value above the more stringent limit set by the Planck collaboration, which includes BAO, $\sum m_i < 0.12~\text{eV}$.

The last three panels of each figure focus on the quark sector, where the first two illustrate correlations among quark mass ratios, while the third presents the relationship between the Dirac \textit{CP} phase and the third quark mixing angle. From the chi-square values associated with the quark sector (see the second-to-last row in Table \ref{results}), it is evident that all eight observables in our numerical analysis fall within the $3\sigma$ range. Notably, in the NO scenario, model II stands out with $\chi^2 \approx 0$, indicating that all observables lie within their $1\sigma$ allowed range.
\section{Unification and proton decay}
\label{sec5}
\subsection{Unification in $SU(5) \times \Gamma_3$ GUT}
It is well known that the minimal non-SUSY $SU(5)$ model has been ruled out for several reasons, including its failure to satisfy the experimental lower bound on the proton lifetime for the decay channel $p \to e^+ \pi^0$ and its lack to achieve successful unification of the three gauge couplings $\alpha_1$, $\alpha_2$, and $\alpha_3$ at high energies \cite{Georgi:1974yf}. In fact, the running of gauge couplings shows that $\alpha_1$ and $\alpha_2$ unify around $10^{12-13}~\text{GeV}$, while $\alpha_2$ and $\alpha_3$ merge at a much higher scale, approximately $10^{16-17}~\text{GeV}$. This mismatch suggests the necessity of additional particles, as postulated in GUTs, to ensure precise unification at high energy scale and increases the prediction of the lifetime for the decay process $p \to e^+ \pi^0$. A viable solution is the introduction of scalar fields below the GUT scale $M_X$, which influence the evolution of gauge couplings through RGEs. Notably, scalar components from the 45-dimensional Higgs representation provide a well-motivated mechanism to achieve unification at experimentally viable energy scales; see Refs \cite{Goto:2023qch, Haba:2024lox, Haba:2024yql} for recent studies on this subject. In our construction, three Higgs fields--$5_H$, $24_H$, and $45_H$--contribute to the RGEs of the gauge couplings. Their decomposition under SM representations is given by
\begin{eqnarray}
5_{H} & = & (3,1,-\frac{1}{3})\oplus(1,2,\frac{1}{2})  \sim T\oplus H\nonumber \\
24_{H} & = & (8,2,\frac{1}{2})\oplus(1,3,0)\oplus(3,2,-\frac{5}{6})\oplus
(\bar{3},2,\frac{5}{6})\oplus \Sigma_{1}(1,1,0) \sim \Sigma_{8}\oplus \Sigma_{3}\oplus \Sigma_{(3,2)}\oplus \Sigma
_{(\overline{3},2)}\oplus \Sigma_{1}\nonumber \\
45_{H} & = & (\bar{3},1,\frac{4}{3})\oplus(\bar{3},2,-\frac{7}{6}%
)\oplus(3,3,-\frac{1}{3})\oplus(6,1,-\frac{1}{3})\oplus(8,2,\frac{1}{2})\oplus(3,1,-\frac{1}{3})\oplus(1,2,\frac{1}{2})\\
&  \sim & \phi_{1}\oplus \phi_{2}\oplus \phi_{3}\oplus \phi_{4}\oplus \phi_{5}\oplus
T^{\prime}\oplus H^{\prime}\nonumber
\end{eqnarray}
In our analysis, we assume that the masses of the heavy $SU(5)$ gauge bosons, $X$ and $Y$, are at the unification scale $M_X = M_{\text{GUT}}$, where the SM gauge couplings unify. At this scale, the gauge coupling constants satisfy the GCU condition $\alpha_3 (M_X) = \alpha_2 (M_X) = \alpha_1 (M_X) = \alpha_X (M_X)$, ensuring their convergence to a single value. The SM gauge couplings, $\alpha_3(\mu)$, $\alpha_2(\mu)$, and $\alpha_1(\mu)$, where $\mu$  is the renormalization scale, are related to the gauge coupling constants $g_s$, $g$, and $g^\prime$ corresponding to $SU(3)_C, SU(2)_L$, and $U(1)_Y$, respectively, as follows
\begin{align}
\alpha_{3}(\mu) = \alpha_{s}(\mu) = \frac{g_{s}(\mu)^{2}}{4\pi}, \quad \alpha_{2}(\mu) = \frac{g(\mu)^{2}}{4\pi}, \quad \alpha_{1}(\mu) = \frac{5}{3} \frac{g^{\prime}(\mu)^{2}}{4\pi}. 
\end{align}
Under the unification assumption, the one-loop RGEs are solved, yielding the following solutions \cite{Goto:2023qch}
\begin{align}
\alpha_X^{-1}(M_X) &= \alpha_3^{-1}(m_Z) - \left( \frac{B_{g_s
}^{SM}}{2\pi} \log \frac{M_X}{m_Z}+ \sum_{\chi} \frac{B_{g_s}
^{\chi}}{2\pi} \log \frac{M_X}{m_{\chi}} \right),
\label{RGE1}\\
\alpha_X^{-1}(M_X) &= \alpha_2^{-1}(m_Z)- \left( \frac
{B_g^{SM}}{2\pi}\log \frac{M_X}{m_Z}+\sum_{\chi}\frac
{B_{g}^{\chi}}{2\pi} \log \frac{M_X}{m_{\chi}}\right),
\label{RGE2}
\\
\alpha_{X}^{-1}(M_X) &= \alpha_{1}^{-1}(m_Z)-\frac{3}{5}\left(
\frac{B_{g^{\prime}}^{SM}}{2\pi} \log \frac{M_X}{m_Z}+\sum_{\chi
} \frac{B_{g^{\prime}}^{\chi}}{2\pi} \log \frac{M_X}{m_{\chi}} \right),
\label{RGE3}
\end{align}
where $m_Z$ is the mass of the $Z$-boson, $\chi$ runs over all relevant scalars, and the coefficients $B_{g_i}^{SM}$ and $B_{g_i}^{\chi}$ denote the beta functions of the three gauge couplings; their corresponding values for the three gauge coupling constants are given in Table \ref{tab4}.
\begin{table}[h]
    \centering
    \renewcommand{\arraystretch}{1.3}
    \setlength{\tabcolsep}{8pt}
    \begin{tabular}{|c|c|c|c|c|c|c|c|c|c|c|c|c|}
        \hline
        $g_i$ & $B_{g_{i}}^{\mathrm{SM}}$ & $B_{g_i}^{H^{\prime}}$ & $B_{g_{i}}^{T}$ & $B_{g_i}^{T^{\prime}}$ & $B_{g_i}^{\phi_1}$ & $B_{g_i}^{\phi_2}$ & $B_{g_i}^{\phi_3}$ & $B_{g_i}^{\phi_4}$ & $B_{g_i}^{\phi_5}$ & $B_{g_i}^{\Sigma_1}$ & $B_{g_i}^{\Sigma_3}$ & $B_{g_i}^{\Sigma_8}$ \\ 
        \hline
        $g_s$ & $-7$ & $0$ & $\frac{1}{6}$ & $\frac{1}{6}$ & $\frac{1}{6}$ & $\frac{1}{3}$ & $\frac{1}{2}$ & $\frac{5}{6}$ & $2$ & $0$ & $0$ & $\frac{1}{2}$ \\ 
        \hline
        $g$ & $-\frac{19}{6}$ & $\frac{1}{6}$ & $0$ & $0$ & $0$ & $\frac{1}{2}$ & $2$ & $0$ & $\frac{4}{3}$ & $0$ & $\frac{1}{3}$ & $0$ \\ 
        \hline
        $g^{\prime}$ & $\frac{41}{6}$ & $\frac{1}{6}$ & $\frac{1}{9}$ & $\frac{1}{9}$ & $\frac{16}{9}$ & $\frac{49}{18}$ & $\frac{1}{3}$ & $\frac{2}{9}$ & $\frac{4}{3}$ & $0$ & $0$ & $0$ \\ 
        \hline
    \end{tabular}
    \caption{Beta functions $B_{g_i}^{SM}$ and $B_{g_{i}}^{\chi}$ for the three gauge coupling constants.}
    \label{tab4}
\end{table}
The relevant SM input parameters used in our analysis are the gauge boson mass $m_Z = 91.1876~\text{GeV}$, the gauge couplings $\alpha_s(m_Z)=0.1193 \pm 0.0016$, $\alpha^{-1}(m_Z)=127.906 \pm 0.019$, and the weak mixing angle $\sin^2 \theta_W(m_Z) = 0.23126 \pm 0.00005$ \cite{PDG}.
The mass splitting among the $45_H$ scalar components plays a crucial role in the RG running of gauge couplings, potentially leading to improved unification. In this context, we investigate different scenarios where certain components remain significantly lighter than the unification scale $M_X$. Their presence affects the RG equations at energy scales above their masses, thereby altering the GCU pattern. Specifically, we explore configurations that achieve GCU by considering cases where one, two, or three scalar components lie below $M_X$ and assess their respective effects on the unification process.\\

\textbf{Scenario I: One scalar component lighter than $M_X$}

\begin{figure}[ptbh]
\begin{center}
\includegraphics[scale=0.4]{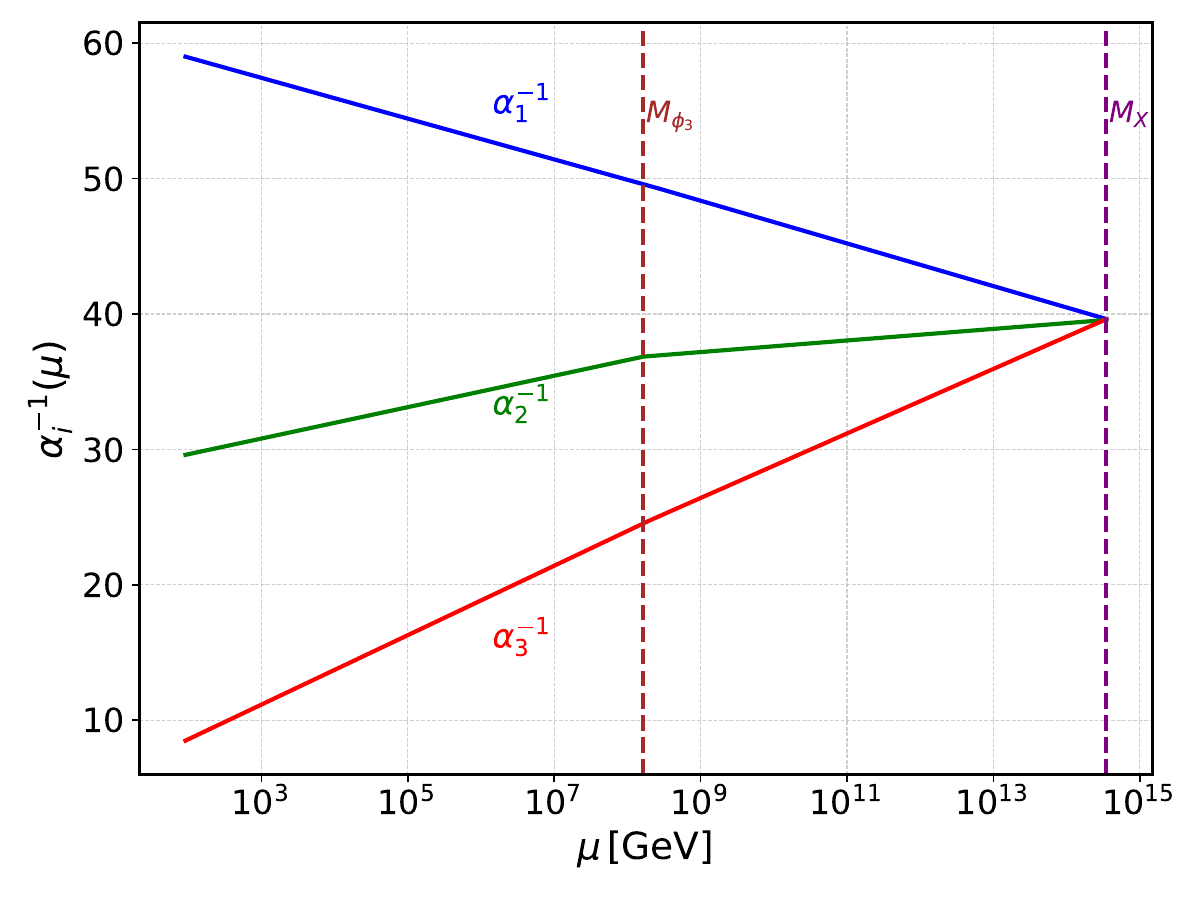} 
\includegraphics[scale=0.4]{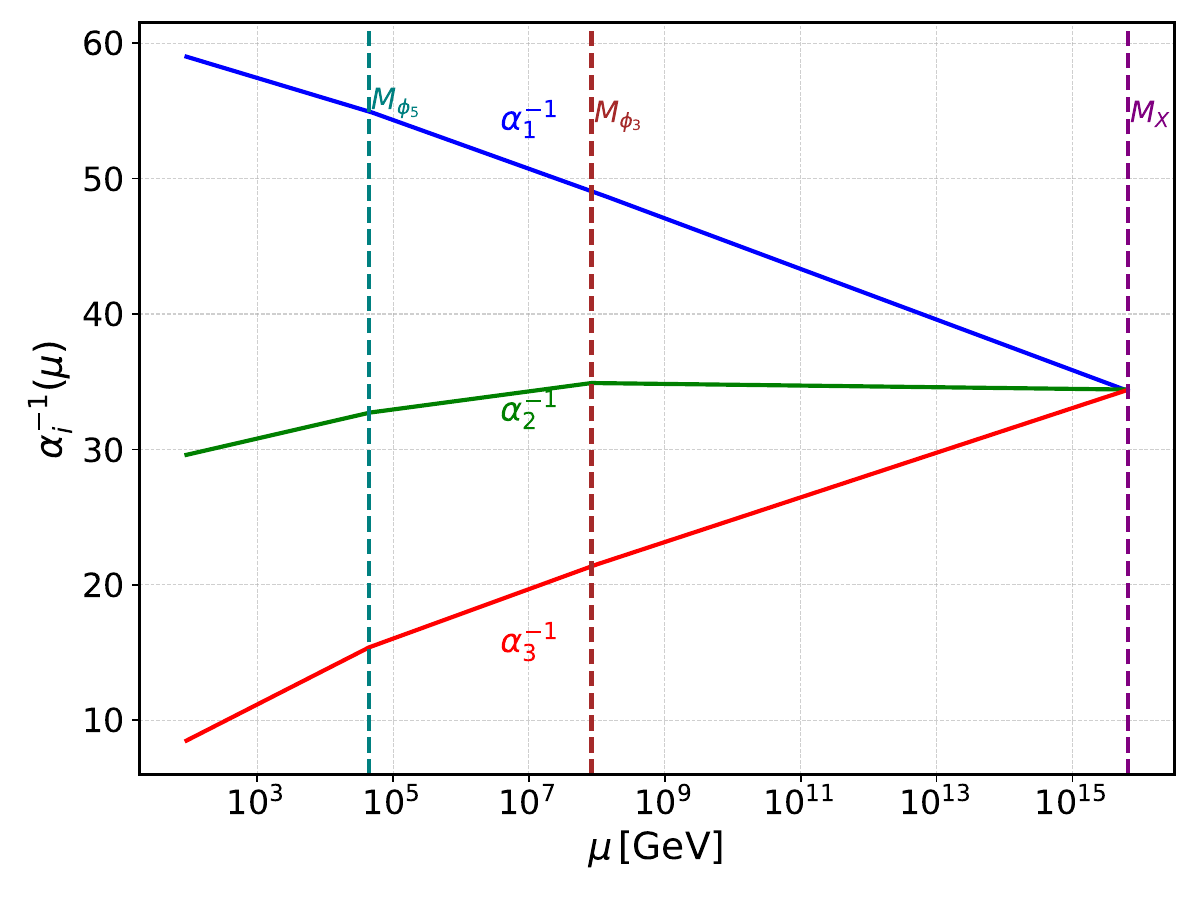}
\includegraphics[scale=0.4]{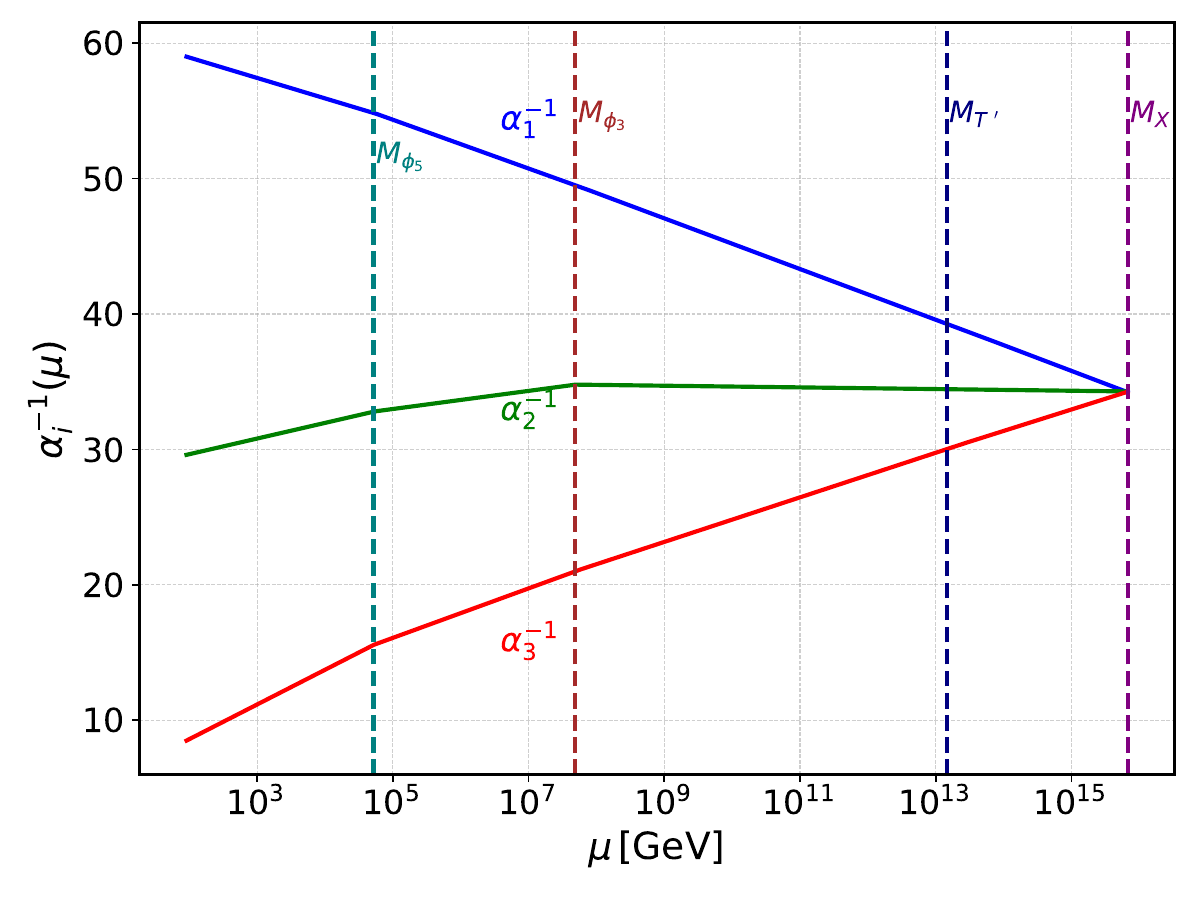}
\end{center}
\par
\vspace{-5mm}
\caption{Renormalization group evolution of gauge coupling constants for three scenarios. Top left panel: Single light scalar $\phi_3$ below $M_X$. Top right panel: Two light scalar components, $\phi_3$ and $\phi_5$, contributing to the running. Bottom panel: Three light scalar components, $\phi_3$, $\phi_5$ and $T^{\prime}$, affecting the gauge coupling unification.}
\label{fig456}
\end{figure}

In the case where a single scalar component of $45_H$ remains lighter than $M_X$, the GCU condition $\alpha_3 (M_X) = \alpha_2 (M_X) = \alpha_1 (M_X) = \alpha_X (M_X)$, as given by Eqs. \ref{RGE1}–\ref{RGE3}, can be fulfilled only if this component is specifically the triplet $\phi_3$. This is due to its significant contribution to the beta function coefficients, which is essential for unification. 
In this context, we find that GCU can occur at a scale $M_X \sim \mathcal{O}(10^{14}~\text{GeV})$ when the mass of $\phi_3$ is approximately $m_{\phi_3} \sim \mathcal{O}(10^{8}~\text{GeV})$. These results are consistent with the findings in Ref. \cite{Haba:2024lox}. Table \ref{tab:masses} presents specific scalar mass configurations for this scenario, while the corresponding gauge coupling evolution is illustrated in the top-left panel of Figure \ref{fig456}.
\begin{table}[h]
    \centering
    \renewcommand{\arraystretch}{1.3}
    \setlength{\tabcolsep}{12pt}
    \begin{tabular}{|l|l|l|l|}
        \hline
        $M_{X}$ & $M_{H^{\prime}},M_{T},M_{T^{\prime}}$ & $M_{\phi_{3}}$ & $M_{\phi_{1}},M_{\phi_{2}},M_{\phi_{4}},M_{\phi_{5}}$ \\ 
        \hline
        $3.49\times10^{14}$ & $3.49\times10^{14}$ & $1.64\times10^{8}$ & $3.49\times10^{14}$ \\ 
        \hline
        $3.70\times10^{14}$ & $3.70\times10^{14}$ & $1.62\times10^{8}$ & $3.70\times10^{14}$ \\ 
        \hline
    \end{tabular}
    \caption{Mass spectrum of scalar components for successful GCU at $M_X \sim \mathcal{O}(10^{14}~\text{GeV})$}
    \label{tab:masses}
\end{table}
This scenario, however, is subject to stringent constraints from proton decay experiments. Specifically, the unification scale $M_X$ is strongly constrained due to contributions from the exchange of the heavy $SU(5)$ gauge bosons $X$ and $Y$, which generate baryon-number-violating dimension-six operators. These operators, in turn, lead to fast proton decay, making the scenario problematic \cite{Nath:2006ut}. 
Moreover, current experimental bounds on the proton lifetime, $\tau_p(p\rightarrow e^+\pi^0) > 2.4 \times 10^{34}~\text{years}$, impose a lower limit of $M_X \gtrsim 5 \times 10^{15}~\text{GeV}$ \cite{Super-Kamiokande:2020wjk}. As a result, this scenario is ruled out by the existing proton decay constraints.
\newline

\textbf{Scenario II: Two scalar components lighter than $M_X$}

To suppress rapid proton decay mediated by the exchange of $X$ and $Y$ gauge bosons, it is essential to significantly raise the unification scale $M_X$. This can be achieved by considering two scalar components acquiring masses below $M_X$. In this scenario, we find that the GCU condition can be satisfied when the scalar triplet $\phi_3$ and the scalar octet $\phi_5$ remain below $M_X$, thereby enabling unification at a scale of $\mathcal{O}(10^{15}~\text{GeV})$. Our analysis shows that the required mass values for unification are $M_{\phi_3} \sim \mathcal{O}(10^7~\text{GeV})$ for the color triplet and $M_{\phi_5} \sim \mathcal{O}(10^4~\text{GeV})$ for the color octet. As an illustration, Table \ref{tab:masses2} provides a summary of three representative cases.
\begin{table}[H]
    \centering
    \renewcommand{\arraystretch}{1.3}
    \setlength{\tabcolsep}{10pt} 
    \begin{tabular}{| c | c | c | c | c | c |} 
        \hline
        $M_X$ & $M_{H^{\prime}},M_{T^{\prime}},M_T$ & $M_{\phi_1},M_{\phi_2},M_{\phi_4}$ & $M_{\phi_3}$ & $M_{\phi_5}$ & $M_{\Sigma_{8}},M_{\Sigma_{3}}$ \\ 
        \hline
        $6.47\times10^{15}$ & $6.47\times10^{15}$ & $6.47\times10^{15}$ & $8.31\times10^7$ & $4.42\times10^4$ & $6.47\times10^{15}$ \\ 
        \hline
        $7.34\times10^{15}$ & $7.34\times10^{15}$ & $7.34\times10^{15}$ & $7.03\times10^7$ & $1.42\times10^4$ & $7.34\times10^{15}$ \\ 
        \hline
        $7.89\times10^{15}$ & $7.89\times10^{15}$ & $7.89\times10^{15}$ & $7.06\times10^7$ & $9.13\times10^3$ & $7.89\times10^{15}$ \\ 
        \hline
    \end{tabular}
    \caption{Mass spectrum of scalar components for successful GCU at $M_X \sim \mathcal{O}(10^{15}~\text{GeV})$.}
    \label{tab:masses2}
\end{table}
Clearly, the mass hierarchy of the scalar components shows that the GCU is successfully achieved when the GUT scale is raised to approximately $M_X \sim 7 \times 10^{15}~\text{GeV}$. The effect of the scalar mass spectra on the evolution of the gauge couplings is shown in the top right panel of Figure \ref{fig456} for one of the cases presented in Table \ref{tab:masses2}. The increase in $M_X$ ensures compatibility with experimental constraints on proton lifetime, as the proton decay from $X$ and $Y$ gauge bosons exchange are sufficiently suppressed.
\newline

\textbf{Scenario III: Three scalar components bellow $M_X$}

A scenario involving three light scalar components below $M_X$ offers greater flexibility in achieving GCU. In this case, we use the scalars $T^\prime$, $\phi_3$, and $\phi_5$ from the $45_H$ Higgs. This scenario maintains compatibility with the experimental constraint $M_X > 5 \times 10^{15}~\text{GeV}$, necessary to prevent rapid proton decay.
\begin{table}[h]
    \centering
    \renewcommand{\arraystretch}{1.3}
    \setlength{\tabcolsep}{10pt}
    \begin{tabular}{| l | l | l | l | l | l | l |}
        \hline
        $M_{X}$ & $M_{H^{\prime}},M_{T}$ & $M_{T^{\prime}}$ & $M_{\phi_{1}},M_{\phi_{2}},M_{\phi_{4}}$ & $M_{\phi_{3}}$ & $M_{\phi_{5}}$ & $M_{\Sigma_{8}},M_{\Sigma_{3}}$ \\ 
        \hline
        $6.72\times10^{15}$ & $6.72\times10^{15}$ & $1.51\times10^{13}$ & $6.72\times10^{15}$ & $4.84\times10^{7}$ & $5.46\times10^{4}$ & $6.72\times10^{15}$ \\ 
        \hline
        $7.31\times10^{15}$ & $7.31\times10^{15}$ & $1.33\times10^{13}$ & $7.31\times10^{15}$ & $5.17\times10^{7}$ & $3.78\times10^{4}$ & $7.31\times10^{15}$ \\ 
        \hline
    \end{tabular}
    \caption{Mass spectrum of scalar components for successful GCU at $M_X \sim \mathcal{O}(10^{15}~\text{GeV})$.}
    \label{tab:masses3}
\end{table}
In Table \ref{tab:masses3}, we present three examples of allowed mass spectra that achieve GCU. Furthermore, one example of gauge coupling unification is illustrated in the bottom panel of Figure \ref{fig456}, where successful unification at the GUT scale is clearly demonstrated.
\subsection{Proton decay}
\label{protondecay}
In $SU(5)$ GUT with a $45_H$ Higgs field, proton decay arises from two primary sources: the exchange of the heavy gauge bosons $X$ and $Y$, and the exchange of colored Higgs triplets. The relevant scalar triplets include $T \sim (3,1,-\frac{1}{3}) \subset 5_H$, $T' \sim (3,3,-\frac{1}{3})$, $\phi_3 \sim (3,3,-\frac{1}{3})$, and $\phi_1 \sim (\bar{3},1,\frac{4}{3}) \subset 45_H$. In both cases, integrating out these multiplets generates effective dimension-six operators that mediate proton decay. For a detailed review on this topic, see Refs. \cite{Nath:2006ut, Dorsner:2012nq}. As shown above, the gauge boson-mediated proton decay processes are suppressed due to the high unification scale. On the other hand, the Higgs triplet-mediated proton decays require the masses of the colored triplets to be sufficiently large to be consistent with the present experimental limits \cite{Super-Kamiokande:2020wjk}. However, achieving GCU while satisfying proton lifetime constraints becomes unrealizable if all color triplet components of the $45_H$ Higgs field have a degenerate mass. In fact, it is found that to avoid rapid proton decay mediated by Higgs triplets, their masses must exceed approximately $3 \times 10^{11}~\text{GeV}$, assuming natural Yukawa couplings \cite{Dorsner:2012uz}. However, if any of these fields is significantly lighter than the GUT scale, their Yukawa couplings must be sufficiently suppressed to remain consistent with the proton lifetime constraints.

\begin{figure}[ptbh]
\begin{center}
\includegraphics[scale=0.4]{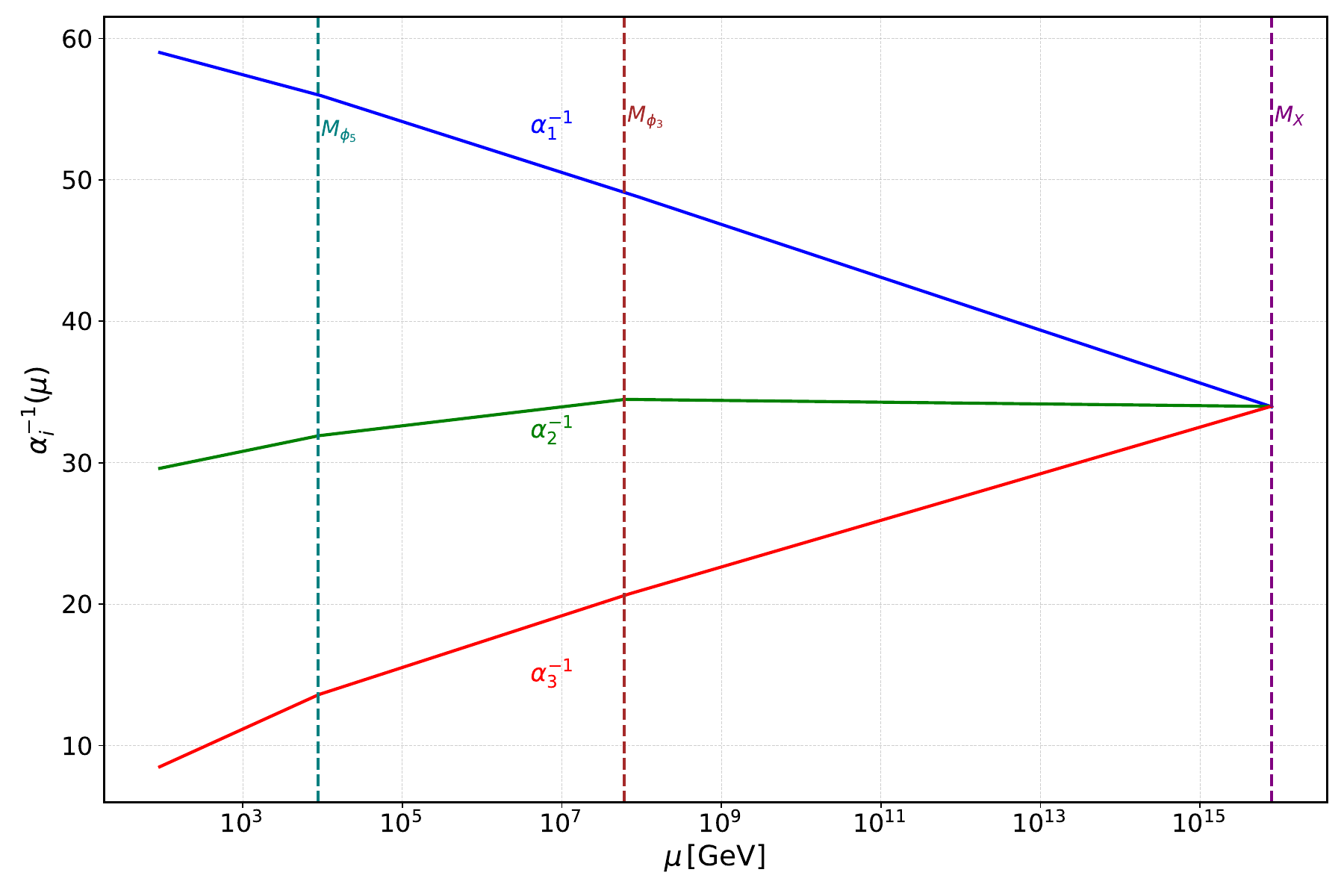}
\end{center}
\par
\vspace{-5mm}
\caption{Evolution of gauge coupling constants in scenario where the scalar triplet $\phi_3$ and the octet $\phi_5$, are set bellow the unification scale $M_X$.}
\label{fig7}
\end{figure}

As previously highlighted, the light scalar $\phi_3$ plays a fundamental role in our framework by enabling unification at a sufficiently high energy scale $M_X$. Thus, we focus here on this scalar triplet, keeping in mind that its mass cannot be arbitrarily light as it must be constrained to comply with experimental proton lifetime limits. To address the tension between these limits and GUT constraints, we propose a scenario in which the masses of the $X$ and $Y$ gauge bosons, along with $T$, $T^{\prime}$ and $\phi_1$ scalar triplets, are set at the scale $\mathcal{O}(10^{15}~\text{GeV})$. Moreover, we impose a mass hierarchy between $\phi_3$, which is fixed slightly below the GUT scale, and the remaining components of the $45_H$ and the $24_H$ fields that affect GCU but do not participate in proton decays process, namely $\{\phi_2,~\phi_4,~\phi_5\} \in 45_H$, and $\{\Sigma_3,~\Sigma_8 \}\in 24_H$. This allows for greater flexibility in assigning their masses while maintaining consistency with unification constraints. For instance, GCU can be achieved with the following mass spectrum
\begin{equation}
    M_{\phi_2} = M_{\phi_4} = M_{\Sigma_3} = M_{\Sigma_8} = 7.916 \times 10^{15}~\text{GeV}, \quad M_{\phi_5} = 8.757 \times 10^3~\text{GeV}, \quad M_{\phi_3} = 6.076 \times 10^7~\text{GeV}.
    \label{mphi3}
\end{equation}
The first equation ensures that the masses of $\phi_2$, $\phi_4$, $\Sigma_3$ and $\Sigma_8$ remain at the GUT scale $M_X$ without disrupting unification. The second is fixed in a way to comply with collider constraints requiring $M_{\phi_5} > 3.1~\text{TeV}$ for natural Yukawa couplings \cite{CMS:2015xau}. Meanwhile, $\phi_3$ in the last equation acquires a mass below the GUT scale playing a crucial role in achieving GCU. We show in Figure \ref{fig7} the RG evolution of the gauge couplings for this mass spectrum, illustrating their successful convergence at the unification scale.
\newline 
We now focus on proton decay mediated by the $\phi_3$ component with its mass given in Eq. \ref{mphi3}, and examine the proton lifetime in the decay process $p \rightarrow e^+ \pi^0$. The corresponding dimension-6 operator, which governs this decay, arises from the following Yukawa interactions
\begin{equation}
    (Y_{\overline{45}})_{ij} Q^T L \phi_3^*, \quad (Y_{45})_{ij} Q^T Q \phi_3.
\end{equation}
Consequently, the proton lifetime can be estimated as
\begin{equation}
    \tau_{p} \sim \frac{M_{\phi_3}^4}{|(Y_{\overline{45}})_{11} (Y_{45})_{11}|^2 m_p^5}.
    \label{lifetime}
\end{equation}
where $m_p \approx 0.94~\text{GeV}$ is the proton mass. In the context of model I with NO, the Yukawa coupling $(Y_{45})_{11}$ is expressed as $(Y_{\overline{45}})_{11} = b_1^{\prime} Y_{3,2}
^{(2)}$. Using the parameter values $|Y_{3,2}^{(2)}|=0.1769$, $(b_{1}^{\prime}/b_{3})=0.5024$, and $b_{3}\upsilon = 57.8613~\text{GeV}$ from Table \ref{results}, we obtain $|(Y_{45})_{11}| = 2.0904 \times 10^{-2}$. Substituting this value into the relation \ref{lifetime} and imposing the current (future) experimental bound from SK [Hyper-Kamiokande (HK)] $\tau_p(p \rightarrow e^+ \pi^0) > 2.4 \times 10^{34}$ years \cite{Super-Kamiokande:2020wjk} ($\tau_p(p \rightarrow e^+ \pi^0) > 7.8 \times 10^{34}$ years \cite{Hyper-Kamiokande:2018ofw}), we find that ensuring the stability of the proton requires $|Y_{{45}})_{11}|\lesssim 1.92 \times 10^{-16}$ ($|Y_{{45}})_{11}| \lesssim 1.06 \times 10^{-16}$). This stringent upper bounds indicate that the contribution of the Yukawa term $Y_{45} 10_{F_i}10_{F_i}45_H$ to the up-quark mass matrix, which we did not consider in our models, is negligibly small. Applying this constraints for the other models using the best-fit values of the parameters in Table \ref{results}, we obtain as well highly suppressed Yukawa couplings where the upper limits are as follows
for $|(Y_{45})_{11}|$:
\begin{eqnarray}
    \text{Model II (NO):} \quad 
    |(Y_{45})_{11}|\lesssim 2.17 \times 10^{-15}~\text{for SK}, \quad \text{and} \quad |(Y_{45})_{11}|\lesssim 1.2 \times 10^{-15} ~\text{for HK}.  \\
    \text{Model II (IO):} \quad 
    |(Y_{45})_{11}|\lesssim 3.95 \times 10^{-15}~\text{for SK}, \quad \text{and} \quad |(Y_{45})_{11}|\lesssim 2.19 \times 10^{-15}~\text{for HK}.  \nonumber
\end{eqnarray}
Thus, the absence of the coupling $Y_{45} 10_{F_i}10_{F_i}45_H$ in our benchmark models is thoroughly justified. 

Finally, we note that beyond the specific assumptions imposed in our benchmark models, additional theoretical mechanisms may be necessary to ensure a more fundamental suppression of this coupling. A particular way in the context of $A_4$ modular invariance, is to assign modular weights for the $45_H$ such that polyharmonic Maa{\ss} forms $Y_r^{k_Y}$--which could generate the operator $Y_r^{k_Y} 10_{F_i} 10_{F_i} 45_H$--are forbidden. This can be enforced by requiring $k_Y \neq k_{10_i} + k_{10_j} + k_{45_{H}}$, ensuring that no polyharmonic Maa{\ss} forms contribute to this interaction. As a result, proton decay mediated by the triplet scalar component $\phi_3$, through the interactions $Y_{45} 10_{F_{i}}10_{F_{i}}45_{H}$ and $Y_{\overline{45}} 10_{F_{i}}5_{F_{i}} 45_{H}^{\ast}$ would be completely absent.
\section{Summary and conclusions}
\label{sec6}
This work introduces nonholomorphic $A_4$ modular symmetry into $SU(5)$ GUT for the first time, providing a unified framework that addresses flavor puzzle and ensures GCU. Specifically, we have employed polyharmonic Maa{\ss} forms of level $N = 3$ and developed two benchmark models which differ from each other by modular weight and $A_4$ charge assignments. After providing the corresponding fermion Yukawa matrices, a numerical scan over the parameter space of each model, including the real and imaginary parts of the modulus $\tau$, was performed to minimize the total chi-square function $\chi_{\text{total}}^2$ and optimize agreement with experimental data. Our analysis demonstrates excellent consistency with experimental results across both the lepton and quark sectors, achieving $\chi_{\text{total}}^2 < 10$, as shown in Table \ref{results}. The results indicate that model I favors only the NO mass spectrum, while model II accommodates both mass orderings.

For neutrino observables, all models yield predictions consistent with the $3\sigma$ ranges from NuFit 6.0, with the exception of a single observable in model II. Specifically, in the NO scenario, the {\it CP}-violating phase $\delta_{CP}^{l}$ lies outside the $3\sigma$ range, while in the IO case, $\sin^{2}\theta_{23}$ exceeds its allowed bound. The atmospheric mixing angle $\theta_{23}^l$ is predicted to be in the higher octant for NO in model I and IO in model II, while it lies in the lower octant for NO in model II. Additionally, the two Majorana {\it CP} phases, $\alpha_{21}$ and $\alpha_{31}$, significantly deviate from the {\it CP}-conserving values of $0$ and $\pi$, with best-fit values clustering around $\pi/2$. For predictions on the absolute neutrino mass scale, the mass parameters $m_{\beta \beta}$, $\sum m_{i}$, and $m_{\beta}$ align with current experimental constraints and provide insights into upcoming detection capabilities. Regarding $m_{\beta \beta}$, the predicted values for all models are consistent with the current most stringent constraint from the KamLAND-Zen collaboration. Similarly, for $m_\beta$, all models yield best-fit values that remain below both current and future sensitivities of the KATRIN experiment. Meanwhile, the predicted values could be tested by the upcoming Project 8 experiment. The sum of neutrino masses for the three models satisfies the limit provided by the latest Planck CMB measurements (TT, TE, EE + lowE + lensing) $\sum m_i < 0.24~\text{eV}$. 
For the charged lepton mass ratios, they are in agreement with their $1\sigma$ level for both models in the NO case, while for model II with IO, they are within their $3\sigma$ range. For the quark observables, we found that they align with data at the GUT scale within $3\sigma$ level. Notably, in the case of model II with NO, all the observables are within the $1\sigma$ level with $\chi^{2}\approx 0$.

The scalar sectors of our benchmark models include a 45-dimensional Higgs field, which is essential for resolving the mass hierarchy between charged leptons and down-type quarks. Additionally, it plays a significant role in the RG evolution of gauge couplings, while some of its components mediate proton decay processes. 
We have examined various scenarios that achieve GCU, with the representative case in Eq. \ref{mphi3} standing out as a particularly successful example. This scenario realizes GCU at $M_X \approx 7.9 \times 10^{15}\text{GeV}$ while remaining consistent with experimental limits on proton decay processes mediated be the heavy gauge bosons $X$ and $Y$. Specifically, a color-triplet scalar $\phi_3$ with $M_{\phi_3} \approx 6 \times 10^7\text{GeV}$ and a color-octet scalar $\phi_5$ with $M_{\phi_5} \approx 8.7 \times 10^3~\text{GeV}$--consistent with the collider constraint $M_{\phi_5} > 3.1~\text{TeV}$--play key roles in achieving unification. On the other hand, proton decay via the scalar triplets $T \in 5_H$ and $T', \phi_1 \in 45_H$ is suppressed due to their GUT-scale masses, while decay via $\phi_3 \in 45_H$ is further suppressed by stringent Yukawa constraints, $|Y_{45})_{11}| \lesssim 10^{-15}$.

This study highlights the viability of nonholomorphic modular flavor symmetries as a bridge between GUT frameworks and observed flavor structures, offering a minimalist alternative to supersymmetric approaches. Future investigations may extend this approach by exploring other modular groups, such as $S_3$, $S_4$ and $A_5$ or incorporating additional Higgs representations to further refine mass predictions.
\section*{Acknowledgements}
The work of M.A.L. and S.N. is supported by the United Arab Emirates University (UAEU) under UPAR Grant No. 12S093.
\appendix
\section{Finite modular group of level 3}
\label{app1}
The finite modular group $\Gamma_3$ is isomorphic to the non-Abelian group $A_4$, which represents the symmetry group of a tetrahedron and consists of the even permutations of four objects. This group is generated by two elements, $S$ and $T$, satisfying the relations $S^2 = (ST)^3 = T^3 = 1$. The order of $A_4$ is given by $4!/2 = 12$, and its elements are distributed among four conjugacy classes
\begin{equation}  
    C_1 = \{ e \}, \quad C_3 = \{S, TST^2, T^2ST \}, \quad C_4 = \{T, ST, TS, STS \}, \quad C_{4'} = \{T^2, ST^2, T^2S, ST^2S \}.
\end{equation}
Since the $A_4$ group has four conjugacy classes, it must also have four irreducible representations. These consist of three one-dimensional representations\footnote{The three one-dimensional representations can be distinguished by their basis characters as  
$\mathbf{1_{(1,1)}}, \mathbf{1_{(1,\omega)}}, \mathbf{1_{(1,\omega^2)}}$, where the entries $(x, y)$ denote the characters of the $S$ and $T$ generators, respectively. For further details on this character notation, see Refs.~\cite{AhlLaamara:2016epq, Ouahid:2018gpg}.}, $\mathbf{1}, \mathbf{1'}, \mathbf{1''}$, and a single three-dimensional representation, $\mathbf{3}$. The representation matrices for the $A_4$ generators in the T-diagonal basis are of the following form for the four different representations
\begin{eqnarray}
    \mathbf{1} &:& S = 1, \quad  T = 1 \nonumber \\
    \mathbf{1'} &:& S = 1, \quad  T = \omega \nonumber \\
    \mathbf{1''} &:& S = 1, \quad  T = \omega^2  \\
    \mathbf{3} &:& S = \begin{pmatrix}
        -1 && 2 && 2 \\
         2 && -1 && 2 \\
         2 && 2 && -1
    \end{pmatrix}, \quad  T = \begin{pmatrix}
        1 && 0 && 0 \\
        0 && \omega && 0 \\
        0 && 0 && \omega^2
    \end{pmatrix}. \nonumber
\end{eqnarray}
with $\omega$ being the cube root of unity: $\omega = e^{2\pi i /3}$. The tensor product rules for the irreducible representations of the $A_4$ group
\begin{eqnarray}
    \mathbf{1} \otimes \mathbf{r} = \mathbf{r}, \quad  \mathbf{1'} \otimes \mathbf{1'} = \mathbf{1''}, \quad  \mathbf{1'} \otimes \mathbf{1''} = \mathbf{1}, \quad  \mathbf{1''} \otimes \mathbf{1''} = \mathbf{1'}, \quad  \mathbf{3} \otimes \mathbf{3} = \mathbf{1} \oplus \mathbf{1'} \oplus \mathbf{1''} \oplus \mathbf{3_S} \oplus \mathbf{3_A},
    \label{tpr}
\end{eqnarray}
where $\mathbf{r}$ denotes any irreducible representation of $A_4$, while $\mathbf{3_S}$ and $\mathbf{3_A}$ stands for the symmetric and the antisymmetric contractions, respectively. The contraction rules for two generic $A_4$ triplets, $a = (a_1, a_2, a_3)^T$ and $b = (b_1, b_2, b_3)^T$, follow from the given tensor product decomposition where we have
\begin{eqnarray}
    (a \otimes b)|_{\mathbf{1}} &=& a_1 b_1 + a_2 b_3 + a_3 b_2 \nonumber \\
    (a \otimes b)|_{\mathbf{1'}} &=& a_3 b_3 + a_1 b_2 + a_2 b_1 \nonumber \\
    (a \otimes b)|_{\mathbf{1''}} &=& a_2 b_2 + a_1 b_3 + a_3 b_1 \\
    (a \otimes b)|_{\mathbf{3_S}} &=& (2 a_1 b_1 - a_2 b_3 - a_3 b_2, 2 a_3 b_3 - a_1 b_2 - a_2 b_1, 2 a_2 b_2 - a_1 b_3 - a_3 b_1 )^T  \nonumber \\
    (a \otimes b)|_{\mathbf{3_A}} &=& (a_2 b_3 - a_3 b_2, a_1 b_2 - a_2 b_1, a_3 b_1 - a_1 b_3 )^T \nonumber
    \label{dypr}
\end{eqnarray}

We now present the explicit expressions for the polyharmonic Maa{\ss} forms $Y^{(k_Y)}_{\mathbf{r}}$ used in our benchmark models, as listed in Tables \ref{tab1} and \ref{tab2}. Notice that these forms correspond to the known modular forms of level $N$ and weight $k \geq 4$, since modular forms with negative weight do not exist \cite{Qu:2024rns}. Moreover, as previously noted, their weight for $\Gamma_N$ is an even integer, where in our case, with $N = 3$, the corresponding finite modular group is the alternating group $A_4$. For our benchmark model I, we have employed polyharmonic Maa{\ss} forms with weights $k_Y = -2, 0, 2, 4, 6$.
\begin{itemize}
    \item $k_Y = -2$: the weight $k_Y = -2$ polyharmonic Maa{\ss} forms can be organized into a trivial singlet $Y_1^{(-2)}(\tau)$ and a triplet $Y_3^{(-2)}(\tau) = [Y_{3,1}^{(-2)}(\tau), Y_{3,2}^{(-2)}(\tau), Y_{3,3}^{(-2)}(\tau)]$ of $A_4$. Their Fourier expansion is given by \cite{Qu:2024rns}
    \begin{align}
        Y_1^{(-2)}(\tau) &= \frac{y^3}{3} - \frac{15 \Gamma(3,4\pi y)}{4\pi^3 q}
        - \frac{135 \Gamma(3,8\pi y)}{32\pi^3 q^2}
        - \frac{35 \Gamma(3,12\pi y)}{9\pi^3 q^3} + \dots \nonumber \\
        &\quad - \frac{\pi}{12} \frac{\zeta(3)}{\zeta(4)}
        - \frac{15q}{2\pi^3} - \frac{135q^2}{16\pi^3} - \frac{70q^3}{9\pi^3} - \frac{1095q^4}{128\pi^3} - \frac{189q^5}{25\pi^3} + \dots
        \label{y1m2}
    \end{align}
    \begin{align}
        Y_{3,1}^{(-2)}(\tau) &= \frac{y^3}{3} + \frac{21 \Gamma(3,4\pi y)}{16\pi^3 q}
        + \frac{189 \Gamma(3,8\pi y)}{128\pi^3 q^2}
        + \frac{169 \Gamma(3,12\pi y)}{144\pi^3 q^3} + \dots \nonumber \\
        &\quad + \frac{\pi}{40} \frac{\zeta(3)}{\zeta(4)}
        + \frac{21q}{8\pi^3} + \frac{189q^2}{64\pi^3} + \frac{169q^3}{72\pi^3} 
        + \frac{1533q^4}{512\pi^3} + \frac{1323q^5}{500\pi^3} + \dots
    \end{align}
    \begin{align}
        Y_{3,2}^{(-2)}(\tau) &= -\frac{729 q^{1/3}}{16\pi^3} \left( 
        \frac{\Gamma(3,8\pi y/3)}{16q} + \frac{7\Gamma(3,20\pi y/3)}{125q^2} 
        + \frac{65\Gamma(3,32\pi y/3)}{1024q^3} + \dots 
        \right) \nonumber \\
        &\quad - \frac{81q^{1/3}}{16\pi^3} \left( 
        1 + \frac{73q}{64} + \frac{344q^2}{343} + \frac{567q^3}{500} 
        + \frac{20198q^4}{2197} + \frac{4681q^5}{4096} + \dots 
        \right)
    \end{align}
    \begin{align}
        Y_{3,3}^{(-2)}(\tau) &= - \frac{81 q^{2/3}}{32\pi^3} 
        \left(\frac{\Gamma(3,4\pi y/3)}{q} 
        + \frac{73\Gamma(3,16\pi y/3)}{64 q^2}
        + \frac{344\Gamma(3,28\pi y/3)}{343 q^3}  + \dots \right) \nonumber \\
        &\quad - \frac{729 q^{2/3}}{8\pi^3} 
        \left(\frac{1}{16} + \frac{7q}{125} + \frac{65q^2}{1024} + \frac{74q^3}{1331} + \dots \right)
    \end{align}
     \item $k_Y = 0$: the weight $k_Y = 0$ polyharmonic Maa{\ss} forms can be organized into a trivial singlet $Y_1^{(0)}(\tau) = 1$ and an $A_4$ triplet $Y_3^{(0)}(\tau) = [Y_{3,1}^{(0)}(\tau), Y_{3,2}^{(0)}(\tau), Y_{3,3}^{(0)}(\tau)]$ where their q-expansion are given by
    \begin{align}
        Y_{3,1}^{(0)} &= y - 3 \frac{e^{-4\pi y}}{\pi q}
        - 9 \frac{e^{-8\pi y}}{2\pi q^2}
        - \frac{e^{-12\pi y}}{\pi q^3}
        - 21 \frac{e^{-16\pi y}}{4\pi q^4}
        - 18 \frac{e^{-20\pi y}}{5\pi q^5} + \dots \nonumber \\
        &\quad - \frac{9 \log 3}{4\pi} - \frac{3q}{\pi} - \frac{9q^2}{2\pi} - \frac{q^3}{\pi} - \frac{21q^4}{4\pi} - \frac{18q^5}{5\pi} - \frac{3q^6}{2\pi} + \dots
    \end{align}
    \begin{align}
        Y_{3,2}^{(0)} &= \frac{27q^{1/3} e^{\pi y/3}}{\pi} \left( 
        \frac{e^{-3\pi y}}{4q} + \frac{e^{-7\pi y}}{5q^2} 
        + \frac{5e^{-11\pi y}}{16q^3} + \frac{2e^{-15\pi y}}{11q^4} 
        + \frac{2e^{-19\pi y}}{7q^5} + \dots 
        \right) \nonumber \\
        &\quad + \frac{9q^{1/3}}{2\pi} \left( 
        1 + \frac{7q}{4} + \frac{8q^2}{7} + \frac{9q^3}{5} 
        + \frac{14q^4}{13} + \frac{31q^5}{16} + \frac{20q^6}{19} + \dots 
        \right) 
    \end{align}
    \begin{align}
        Y_{3,3}^{(0)} &= \frac{9 q^{2/3} e^{2\pi y/3}}{2\pi} \left( 
        \frac{e^{-2\pi y}}{q} + \frac{7e^{-6\pi y}}{4q^2} 
        + \frac{8e^{-10\pi y}}{7q^3} + \frac{9e^{-14\pi y}}{5q^4} 
        + \frac{14e^{-18\pi y}}{13q^5} + \dots 
        \right) \nonumber \\
        &\quad + \frac{27q^{2/3}}{\pi} \left( 
        \frac{1}{4} + \frac{q}{5} + \frac{5q^2}{16} + \frac{2q^3}{11} 
        + \frac{2q^4}{7} + \frac{9q^5}{17} + \frac{21q^6}{20} + \dots 
        \right)
    \end{align}
     \item $k_Y = 2$: The weight 2 polyharmonic Maa{\ss} forms consist of the modified Eisenstein series $\hat{E}_2(\tau)$, which is a trivial singlet under $A_4$, and the modular form triplets of weight 2 and level 3, denoted as $Y_3^{(2)}(\tau) = (Y_1(\tau), Y_2(\tau), Y_3(\tau))^T$. The modified Eisenstein series is defined as $\hat{E}_2(\tau) = E_2(\tau) - \frac{3}{\pi y}$ where $E_2(\tau)$ is the weight 2 Eisenstein series expressed as $E_2(\tau) = 1 - 24 \sum_{n=1}^\infty \sigma_1(n) q^n$ with $\sigma_1(n) = \sum_{d|n}d$ is the sum of the divisors of $n$. The q-expansions of the components of the triplet $Y_3^{(2)}(\tau)$ are given as follows \cite{Feruglio:2017spp}
     \begin{align}
        Y_1(\tau) = 1 + 12q + 36q^2 + ..., \quad Y_2(\tau) = -6q^{1/3} (1 + 7q + 8q^2 + ...), \quad Y_3(\tau) = -18q^{2/3} (1 + 2q + 5q^2 + ...).
     \end{align}
    \item $k_Y = 4$: Weight 4 polyharmonic Maa{\ss} forms correspond to modular forms at level 3. These forms arise from the tensor product $Y_3^{(2)}(\tau) \otimes Y_3^{(2)}(\tau)$, which decomposes into a trivial singlet $Y_1^{(4)}(\tau)$, a nontrivial singlet $Y_{1'}^{(4)}(\tau)$, and an $A_4$ triplet $Y_3^{(4)}(\tau)$, whose explicit expressions are given as follows
    \begin{eqnarray}
        Y_1^{(4)}(\tau) &=& Y_1^2(\tau) + 2 Y_2(\tau) Y_3(\tau) \nonumber \\
        Y_{1'}^{(4)}(\tau) &=& Y_3^2(\tau) + 2 Y_1(\tau) Y_2(\tau) \\
        Y_3^{(4)}(\tau) &=& ( Y_1^2(\tau) - Y_2(\tau) Y_3(\tau) ,  Y_3^2(\tau) - Y_1(\tau) Y_2(\tau),  Y_2^2(\tau) + 2 Y_1(\tau) Y_3(\tau) )^T \nonumber.
    \end{eqnarray}
    \item $k_Y = 6$: Similar to the preceding case, weight 6 polyharmonic Maa{\ss} forms correspond to modular forms at level 3. There are three such forms: a trivial singlet $Y_1^{(6)}(\tau)$ and two $A_4$ triplets denoted as $Y_{3I}^{(6)}(\tau)$ and $Y_{3II}^{(6)}(\tau)$ \cite{Qu:2024rns}. In our benchmark model I, we used only the trivial singlet which is expressed as: $Y_1^{(6)}(\tau) = Y_1^3(\tau) + Y_2^3(\tau) + Y_3^3(\tau) - 3 Y_1(\tau) Y_2(\tau) Y_3(\tau)$. 
\end{itemize}
In our benchmark model II, we have employed polyharmonic Maa{\ss} forms with weights $k_Y = -4, -2, 4, 6$. The explicit expressions of the weights $-2$, $4$ and $6$ polyharmonic Maa{\ss} forms are given in the previous case of model I, while weight $-4$  polyharmonic Maa{\ss} forms consist of a trivial singlet $Y_1^{(-4)}(\tau)$ and an $A_4$ triplet $Y_1^{(-4)}(\tau) = [Y_{3,1}^{(-4)}(\tau), Y_{3,2}^{(-4)}(\tau), Y_{3,3}^{(-4)}(\tau)]$ expressed as \cite{Qu:2024rns}
\begin{align}
    Y_1^{(-4)} &= \frac{y^5}{5} + \frac{63\Gamma(5,4\pi y)}{128\pi^5q} 
    + \frac{2079\Gamma(5,8\pi y)}{4096\pi^5q^2} 
    + \frac{427\Gamma(5,12\pi y)}{864\pi^5q^3} + \dots \nonumber \\
    &\quad + \frac{\pi}{80} \frac{\zeta(5)}{\zeta(6)}
    + \frac{189q}{16\pi^5} + \frac{6237q^2}{512\pi^5} + \dots 
\end{align}
\begin{align}
    Y_{3,1}^{(-4)} &= \frac{y^5}{5} - \frac{549}{3328\pi^5} \left( 
    \frac{\Gamma(5,4\pi y)}{q} + \frac{33\Gamma(5,8\pi y)}{32q^2} 
    + \frac{14641\Gamma(5,12\pi y)}{14823q^3} + \dots 
    \right) \nonumber \\
    &\quad - \frac{3\pi}{728} \frac{\zeta(5)}{\zeta(6)}
    - \frac{1647}{416\pi^5} \left( 
    q + \frac{33q^2}{32} + \frac{14641q^3}{14823} + \dots 
    \right) 
\end{align}
\begin{align}
    Y_{3,2}^{(-4)} &= \frac{72171 q^{1/3}}{212992\pi^5} \left( 
    \frac{\Gamma(5,8\pi y/3)}{q} + \frac{33344\Gamma(5,20\pi y/3)}{34375q^2} 
    + \frac{1025\Gamma(5,32\pi y/3)}{1024q^3} + \dots 
    \right) \nonumber \\
    &\quad + \frac{6561q^{1/3}}{832\pi^5} \left( 
    1 + \frac{1057q}{1024} + \frac{16808q^2}{16807} + \dots 
    \right) 
\end{align}
\begin{align}
    Y_{3,3}^{(-4)} &= \frac{2187 q^{2/3}}{6656\pi^5} 
    \left(\frac{\Gamma(5,4\pi y/3)}{q} 
    + \frac{1057\Gamma(5,16\pi y/3)}{1024 q^2} 
    + \frac{16808\Gamma(5,28\pi y/3)}{16807 q^3} + \dots \right) \nonumber \\
    &\quad + \frac{216513 q^{2/3}}{26624\pi^5} 
    \left(1 + \frac{33344q}{34375} + \frac{1025q^2}{1024} 
    + \frac{1717888q^3}{1771561} + \frac{16808q^4}{16807} + \dots \right)
    \label{y3m4}
\end{align}
\section{Scalar potential and masses of the scalar components of $5_H$, $24_H$ and $45_H$}
\label{app2}
In this appendix, we show that the mass scales of the scalar triplet $\phi_3$ and the scalar octet $\phi_5$, introduced to achieve successful GCU while avoiding rapid proton decay mediated by colored Higgs triplets, are consistent with their corresponding analytical expressions. To ensure completeness, we present the mass spectra of all scalar field components relevant our study, laying the basis to explore whether the scalar potential provides sufficient parametric freedom to accommodate the desired mass splitting between $\phi_3$ and $\phi_5$. 
The most general scalar potential for a non-SUSY $SU(5)$ GUT involving the Higgs representations $5_H$, $24_H$, and $45_H$, takes the form
\begin{eqnarray}
    V(H_5, \Sigma_{24},H_{45}) = V_{24} + V_{5} + V_{45} + V_{24,5} + V_{24,45} + V_{5,45} + V_{24,5,45}
\end{eqnarray}
where $V_{24}$, $V_5$, and $V_{45}$ denote the individual potentials for each Higgs multiplet, and the remaining terms account for interactions between different representations and are given explicitly by
\begin{align}
     V_{24} &= \mu_{24}^2\, \mathrm{tr}(\Sigma_{24}^2) + \chi_{24}\, \mathrm{tr}(\Sigma_{24}^3) + \lambda_{24}^{(1)} \left( \mathrm{tr}(\Sigma_{24}^2) \right)^2 + \lambda_{24}^{(2)}\, \mathrm{tr}(\Sigma_{24}^4), \nonumber \\
	V_{5} &= \mu_{5}^2\, H_5^\dagger H_5 + \lambda_5 (H_5^\dagger H_5)^2, \nonumber \\
	V_{45} &= \mu_{45}^2 (H_{45}^\dagger)^i_{jk}(H_{45})^{jk}_i 
	+ \lambda_{45}^{(1)} \left( (H_{45}^\dagger)^i_{jk}(H_{45})^{jk}_i \right)^2 
	+ \lambda_{45}^{(2)} (H_{45}^\dagger)^i_{jk}(H_{45})^{jk}_l (H_{45}^\dagger)^l_{mn}(H_{45})^{mn}_i \nonumber\\
	&\quad + \lambda_{45}^{(3)} (H_{45}^\dagger)^i_{jk}(H_{45})^{jn}_i (H_{45}^\dagger)^l_{mn}(H_{45})^{mk}_l 
	+ \lambda_{45}^{(4)} (H_{45}^\dagger)^i_{jk}(H_{45})^{jk}_n (H_{45}^\dagger)^l_{mi}(H_{45})^{mn}_l \nonumber\\
	&\quad + \lambda_{45}^{(5)} (H_{45}^\dagger)^i_{jk}(H_{45}^\dagger)^j_{il} (H_{45})^{mk}_n (H_{45})^{nl}_m 
	+ \lambda_{45}^{(6)} (H_{45}^\dagger)^i_{jk}(H_{45}^\dagger)^j_{lm} (H_{45})^{kl}_n (H_{45})^{mn}_i,
	\nonumber\\
	V_{24,5} &= \chi_5\, H_5^\dagger \Sigma_{24} H_5 + a^{(1)} (\mathrm{tr}(\Sigma_{24}^2)) H_5^\dagger H_5 + a^{(2)} H_5^\dagger \Sigma_{24}^2 H_5,
	\nonumber\\
		V_{24,45} &= \chi_{45}^{(1)} (H_{45}^\dagger)^i_{jk} (\Sigma_{24})^l_{\ i} (H_{45})^{jk}_l 
	+ \chi_{45}^{(2)} (H_{45}^\dagger)^i_{jk} (\Sigma_{24})^k_{\ l} (H_{45})^{jl}_i \nonumber\\
	&\quad + b^{(1)} (\mathrm{tr}(\Sigma_{24}^2)) (H_{45}^\dagger)^i_{jk} (H_{45})^{jk}_i 
	+ b^{(2)} (H_{45}^\dagger)^i_{jk} (\Sigma_{24}^2)^l_{\ i} (H_{45})^{jk}_l \nonumber\\
	&\quad + b^{(3)} (H_{45}^\dagger)^i_{jk} (\Sigma_{24}^2)^j_{\ l} (H_{45})^{lk}_i 
	+ b^{(4)} (H_{45}^\dagger)^i_{jk} (\Sigma_{24})^m_{\ i} (\Sigma_{24})^j_{\ l} (H_{45})^{lk}_m \\
	&\quad + b^{(5)} (H_{45}^\dagger)^i_{jk} (\Sigma_{24})^k_{\ m} (\Sigma_{24})^j_{\ l} (H_{45})^{lm}_i 
	+ b^{(6)} (H_{45}^\dagger)^i_{jk} (\Sigma_{24})^j_{\ i} (\Sigma_{24})^m_{\ l} (H_{45})^{lk}_m,
	\nonumber\\
	V_{5,45} &= c^{(1)} (H_{45}^\dagger)^i_{jk} (H_{45})^{jk}_i (H_5^\dagger H_5) 
	+ c^{(2)} (H_5^\dagger)_i (H_{45}^\dagger)^i_{jk} (H_{45})^{jk}_l (H_5)^l \nonumber\\
	&\quad + c^{(3)} (H_5^\dagger)_k (H_{45})^{jk}_i (H_{45}^\dagger)^i_{jl} (H_5)^l 
	+ c^{(4)} (H_{45})^{jk}_i (H_{45})^{il}_j (H_5^\dagger)_k (H_5^\dagger)_l \nonumber\\
	&\quad + c^{(5)} (H_{45}^\dagger)^i_{jk} (H_{45})^{jk}_l (H_{45})^{lm}_i (H_5^\dagger)_m 
	+ c^{(6)} (H_{45}^\dagger)^i_{jk} (H_{45})^{jl}_i (H_{45})^{km}_l (H_5^\dagger)_m + \mathrm{h.c.},
	\nonumber\\
		V_{24,5,45} &= \tilde{\chi}\, (H_5^\dagger)_k (\Sigma_{24})^i_{\ j} (H_{45})^{jk}_i 
	+ d^{(1)} (H_5^\dagger)_k (\Sigma_{24}^2)^i_{\ j} (H_{45})^{jk}_i + d^{(2)} (H_5^\dagger)_l (\Sigma_{24})^l_{\ k} (\Sigma_{24})^i_{\ j} (H_{45})^{jk}_i + \mathrm{h.c.}, \nonumber
\end{align}
with $i,~j,~k,~l,~m,~n$ denoting $SU(5)$ indices. The scalar field $\Sigma_{24}$ acquires a VEV that  breaks the $SU(5)$ gauge symmetry along the  direction
\begin{equation}
	\langle \Sigma_{24} \rangle = \begin{pmatrix}
		2 v_{24} & 0 & 0 & 0 & 0 \\
		0 & 2 v_{24} & 0 & 0 & 0 \\
		0 & 0 & 2 v_{24} & 0 & 0 \\
		0 & 0 & 0 & -3 v_{24} & 0 \\
		0 & 0 & 0 & 0 & -3 v_{24}
	\end{pmatrix},
\end{equation}
For clarity, we use the following parameterization of the scalar potential's free parameters\cite{Goto:2023qch}
\begin{eqnarray}
    \tilde{\mu}_{5}^2 &=& \mu_{5}^2 + \left( 30 a^{(1)} + 6 a^{(2)} \right) v_{24}^2, \quad \tilde{\mu}_{45}^2 = \mu_{45}^2 + \left( 30 b^{(1)} + 6 b^{(2)} + 6 b^{(3)} - \frac{3}{2} b^{(4)} + b^{(5)} \right) v_{24}^2, \nonumber\\
    \tilde{a}^{(2)} &=& a^{(2)} - \frac{\chi_5}{v_{24}}, \quad
    \tilde{b}^{(2)} = b^{(2)} - \frac{\chi_{45}^{(1)}}{v_{24}},
    \quad \tilde{b}^{(3)} = b^{(3)} - \frac{\chi_{45}^{(2)}}{v_{24}}, \quad \tilde{d}^{(1)} = d^{(1)} -\frac{\tilde{\chi}}{v_{24}}.
    \label{scalarmasses2}  
\end{eqnarray}
The squared masses of the scalar components in $5_H$, $25_H$, and $45_H$ representations are then derived from the scalar potential $V(H_5, \Sigma_{24},H_{45})$, which encodes both their self-interactions and cross-interactions. Explicitly, we obtain
\begin{eqnarray}
    M^2_{\Sigma_1} &=& -2 \mu_{24}^2 + \frac{3}{2} \chi_{24} v_{24} ,\quad m^2_{\Sigma_3} = 40 \lambda_{24}^{(2)} v_{24}^2 - \frac{15}{2} \chi_{24} v_{24}, \quad
    M^2_{\Sigma_8} = 10 \lambda_{24}^{(2)} v_{24}^2 + \frac{15}{2} \chi_{24} v_{24}, \nonumber\\
	M^2_{\phi_1} &=& \tilde{\mu}_{45}^2 + \left( -2 \tilde{b}^{(2)} + 3 \tilde{b}^{(3)} - \frac{9}{2} b^{(4)} + 8 b^{(5)} \right) v_{24}^2, \quad
	M^2_{\phi_2} = \tilde{\mu}_{45}^2 + \left( 3 \tilde{b}^{(2)} - 2 \tilde{b}^{(3)} - \frac{9}{2} b^{(4)} + 3 b^{(5)} \right) v_{24}^2, \nonumber\\
	M^2_{\phi_3} &=& \tilde{\mu}_{45}^2 + \left( 3 \tilde{b}^{(2)} + \frac{1}{2} \tilde{b}^{(3)} + 3 b^{(4)} - 7 b^{(5)} \right) v_{24}^2, \quad
	M^2_{\phi_4} = \tilde{\mu}_{45}^2 + \left( -2 \tilde{b}^{(2)} - 2 \tilde{b}^{(3)} + \frac{11}{2} b^{(4)} + 3 b^{(5)} \right) v_{24}^2,
	\nonumber\\
	M^2_{\phi_5} &=& \tilde{\mu}_{45}^2 + \left( -2 \tilde{b}^{(2)} + \frac{1}{2} \tilde{b}^{(3)} + \frac{1}{2} b^{(4)} - 7 b^{(5)} \right) v_{24}^2,
	\label{scalarmasses1}	
\end{eqnarray}
The scalar doublets $H$ and $H'$, as well as the  triplets $T$ and $T'$, have the same quantum numbers under the SM gauge group and can therefore mix. Their physical mass eigenstates $h$, $h'$, $h_C$, and $h'_C$ are defined through mixing angles $\theta_H$ and $\theta_S$, and the phases $\delta_H$ and $\delta_S$, expressed as 
\begin{equation}
\begin{pmatrix}
	h \\
	h'
\end{pmatrix}
=
\begin{pmatrix}
	\cos\theta_H & e^{-i \delta_H} \sin\theta_H \\
	- e^{i \delta_H} \sin\theta_H & \cos\theta_H
\end{pmatrix}
\begin{pmatrix}
	H \\
	H'
\end{pmatrix},
\quad
\begin{pmatrix}
	h_C \\
	h'_C
\end{pmatrix}
=
\begin{pmatrix}
	\cos\theta_S & e^{-i \delta_S} \sin\theta_S \\
	- e^{i \delta_S} \sin\theta_S & \cos\theta_S
\end{pmatrix}
\begin{pmatrix}
	T \\
	T'
\end{pmatrix}.
\end{equation}
The corresponding mass-squared matrices $m^2_{h,h'}$ and $m^2_{h_C,h'_C}$ can be rewritten in a compact form in terms of three parameters each
\begin{equation}
	m^2_{h,h'} = 
	\begin{pmatrix}
		A & B \\
		B^* & C
	\end{pmatrix}
	\quad , \quad
	m^2_{h_C,h'_C} =
	\begin{pmatrix}
		A' & B' \\
		B'^* & C'
	\end{pmatrix},
\end{equation}
where
\begin{eqnarray}
   A &=& \tilde{\mu}_5^2 + 3 \tilde{a}^{(2)} v_{24}^2, \quad B = -\dfrac{5\sqrt{3}}{2\sqrt{2}} \left( \tilde{d}^{(1)} + 3 d^{(2)} \right) v_{24}^2, \quad A' = \tilde{\mu}_5^2 - 2 \tilde{a}^{(2)} v_{24}^2, \nonumber \\
     B' &=& -\dfrac{5}{\sqrt{2}} \left( \tilde{d}^{(1)} - 2 d^{(2)} \right) v_{24}^2, \quad C = \tilde{\mu}_{45}^2 + \left( \dfrac{7}{4} \tilde{b}^{(2)} + \dfrac{19}{8} \tilde{b}^{(3)} + 8 b^{(4)} + \dfrac{17}{4} b^{(5)} + \dfrac{75}{8} b^{(6)} \right) v_{24}^2, \\
     C' &=& \tilde{\mu}_{45}^2 + \left( \dfrac{1}{2} \tilde{b}^{(2)} - \dfrac{3}{4} \tilde{b}^{(3)} + \dfrac{17}{4} b^{(4)} - 2 b^{(5)} + \dfrac{25}{2} b^{(6)} \right) v_{24}^2. \nonumber
\end{eqnarray}
The resulting mass eigenvalues are given by
\begin{equation}
    M^2_{h,h'} = \frac{A+C}{2} \mp \sqrt{ \left( \frac{A-C}{2} \right)^2 + |B|^2 }, \quad M^2_{h_C,h'_C} = \frac{A'+C'}{2} \pm \sqrt{ \left( \frac{A'-C'}{2} \right)^2 + |B'|^2 }.
    \label{scalarmasses3}
\end{equation}
It is clear that the mass-squared expressions for the scalar components in the $5_H$ and $45_H$ Higgs multiplets, given in Eqs.~\eqref{scalarmasses1} and \eqref{scalarmasses3}, form a solvable system when the physical masses are set to the numerical values specified in subsection~\ref{protondecay}. Specifically\footnote{To simplify our numerical study of gauge coupling unification, we disregarded the mixing effects between $T$ and $T'$ and between $H$ and $H'$.}, $\{T, T^{\prime}, \phi_1\} \sim \mathcal{O}(10^{15}~\text{GeV})$, $M_{\phi_2} = M_{\phi_4} = M_{\Sigma_3} = M_{\Sigma_8} = 7.916 \times 10^{15}~\text{GeV}$, $M_{\phi_5} = 8.757 \times 10^3~\text{GeV}$, and $M_{\phi_3} = 6.076 \times 10^7~\text{GeV}$. This solvability stems from the fact that the number of free parameters in the analytical expressions for the scalar masses exceeds the number of mass constraints. This surplus of variables allows multiple consistent solutions, providing significant flexibility in fitting the spectrum while satisfying physical and phenomenological requirements. For instance, the mass-squared expressions for the scalar fields $m_{\phi_i}^2$, the Higgs doublets $m_{h,h'}^2$, and the color triplets $m_{h_C}, m_{h'_C}^2$ all depend on a common set of independent parameters. Collectively, these yield a system of nine mass constraints, expressed as a function of at least seventeen independent free parameters. Fixing the scalar masses to the benchmark values required for successful GCU leads to an underconstrained system. The excess of parameters over constraints ensures that the system remains solvable, allowing for multiple viable parameter configurations that consistently reproduce the full mass spectrum.

Note that  there are many ways to fix some of the free parameters in the mass-squared expressions. For example, by evaluating the mass splitting between the scalar masses squared of $\phi_3$ and $\phi_5$ using their expressions in Eq.~(\ref{scalarmasses1}), we obtain the simple relation
\begin{equation}
M^2_{\phi_3} - M^2_{\phi_5} = \left(5\tilde{b}^{(2)} + \frac{5}{2}b^{(4)}\right)v_{24}^2
\end{equation}
Employing the benchmark values of the octet and the triplet masses from Eq.~(\ref{mphi3}), the numerical mass splitting is found to be $M^2_{\phi_3} - M^2_{\phi_5} \approx 3.69\times10^{15}~\text{GeV}^2$, which yields the following constraint on the coupling constants
\begin{equation}
5\tilde{b}^{(2)} + \frac{5}{2}b^{(4)} \approx \frac{3.69\times10^{15}\,\text{GeV}^2}{v_{24}^2}.
\end{equation}
Taking the scale of $SU(5)$ symmetry breaking as $\upsilon_{24} \sim 10^{15}~\text{GeV}$ allows one to fix the coupling constants $\tilde{b}^{(2)}$ and $b^{(4)}$ and chose them to be positive. Additional  free parameters can be fixed by minimizing the scalar potential as detailed in Refs.~\cite{Kalyniak:1982pt,Eckert:1983bn} for the $SU(5)$ scalar potential involving $45_H$ Higgs.

\bibliographystyle{JHEP}
\bibliography{bibliography.bib}

\end{document}